\documentclass{SciPost}

\newif\ifsubmissiondraft
\submissiondraftfalse

\usepackage{hyperref}
\hypersetup{
	colorlinks,
	linkcolor={red!50!black},
	citecolor={blue!50!black},
	urlcolor={blue!80!black}
}
\usepackage[bitstream-charter]{mathdesign}
\DeclareSymbolFont{usualmathcal}{OMS}{cmsy}{m}{n}
\DeclareSymbolFontAlphabet{\mathcal}{usualmathcal}

\usepackage{forest}
\usepackage{mathtools}
\usepackage{bm}

\fancypagestyle{SPstyle}{
	\fancyhf{}
	\lhead{\colorbox{scipostblue}{\bf \color{white} ~SciPost Physics }}
	\rhead{{\bf \color{scipostdeepblue} ~Submission }}
	
	\fancyfoot[C]{\textbf{\thepage}}
}
\fancypagestyle{ARXIVstyle}{
	\fancyhf{}
	
	\fancyfoot[C]{\textbf{\thepage}}
}

\newcommand{\cS}{\mathbf{S}}

\newcommand{\btdot}{%
	\begin{forest}
		for tree={grow'=90,parent anchor=center,child anchor=center}
		[]
		\path[fill=black] (.anchor) circle[radius=1.5pt];
	\end{forest}%
}
\newcommand{\bttree}[1]{%
	\begin{forest}
		for tree={grow'=270,parent anchor=center,child anchor=center,l=0cm,inner ysep=0cm,s sep=0.8mm,edge+=thick}
		#1
	\end{forest}%
}

\begin{document}
	\ifsubmissiondraft
	\pagestyle{SPstyle}
	\else
	\makeatletter
	\let\ps@plain\ps@ARXIVstyle
	\makeatother
	\pagestyle{ARXIVstyle}
	\fi
	
	\begin{center}{\Large \textbf{\color{scipostdeepblue}{
					Equilibrium in a Reaction Network of Assemblies
	}}}\end{center}
	
	\begin{center}\textbf{
			Giampaolo Folena\textsuperscript{1} and
			Germ\'an Kruszewski\textsuperscript{2}
	}\end{center}
	
	\begin{center}
		{\bf 1} Institute for Cross-disciplinary Physics and Complex Systems IFISC (CSIC-UIB), Campus Universitat Illes Balears, 07122 Palma de Mallorca, Spain
		\\
		{\bf 2} Computational Linguistics and Linguistic Theory Lab, Universitat Pompeu Fabra, Barcelona, Spain
	\end{center}
	
	\section*{\color{scipostdeepblue}{Abstract}}
	\textbf{
		We study a mean-field reaction network whose species are assemblies built from identical atoms by reversible coagulation and fragmentation. Each assembly is an ordered binary tree, so the number of species of a given length grows combinatorially,
		as the Catalan numbers. The model nonetheless admits an explicit equilibrium and tractable stochastic dynamics. A finite volume $V$ sets a crossover length $l_c \sim \ln V$ that splits the equilibrium into two sectors. Below $l_c$ each assembly occurs in many copies and the rank-frequency distribution is Zipf-like; above $l_c$ individual species are rare and fluctuation-dominated. The statistical weight of the rare sector decays slowly with volume, controlling the finite-size scaling of diversity, Shannon entropy, and other assembly-weighted observables.
		The equilibrium also admits a transparent grand-canonical description in terms of a bond energy and an atomic chemical potential. Together these results make the model a controlled neutral baseline against which selection and driving in richer assembly networks can be measured.
	}
	
	\vspace{\baselineskip}
	
	\ifsubmissiondraft
	\noindent\textcolor{white!90!black}{%
		\fbox{\parbox{0.975\linewidth}{%
				\textcolor{white!40!black}{\begin{tabular}{lr}%
						\begin{minipage}{0.6\textwidth}%
							{\small Copyright attribution to authors. \newline
								This work is a submission to SciPost Physics. \newline
								License information to appear upon publication. \newline
								Publication information to appear upon publication.}
						\end{minipage} & \begin{minipage}{0.4\textwidth}
							{\small Received Date \newline Accepted Date \newline Published Date}%
						\end{minipage}
				\end{tabular}}
		}}
	}
	
	\linenumbers
	\fi
	\vspace{10pt}
	\noindent\rule{\textwidth}{1pt}
	\tableofcontents
	\noindent\rule{\textwidth}{1pt}
	\vspace{10pt}
	
	\section{Introduction}
	
	The question of how many distinct molecules a set of atoms can form is among the oldest in mathematical chemistry. In 1875 Arthur Cayley posed it for the alkanes $\mathrm{C}_n\mathrm{H}_{2n+2}$, recognizing that counting structural isomers is equivalent to counting unlabeled bounded-valence trees, and introducing generating functions to do so~\cite{Cayley1875}. This founding insight, later generalized by P\'olya's enumeration theorem~\cite{Polya1937}, shows that the space of possible structures explodes combinatorially with size: there are $75$ alkane isomers of $\mathrm{C}_{10}\mathrm{H}_{22}$ but already $366{,}319$ of $\mathrm{C}_{20}\mathrm{H}_{42}$~\cite{oeisA000602}. Turning such recursive structural definitions into sharp asymptotic counts became a driving problem of analytic combinatorics~\cite{FlajoletSedgewick2009}.
	
	Yet enumeration is a \emph{static} statement: it counts how many structures exist at each size, not how a closed, reacting pool of atoms populates them. Where is the dynamics? This is the domain of chemical reaction networks, the standard framework for relating microscopic stochasticity, macroscopic kinetics, and thermodynamic dissipation~\cite{Kurtz1972,Risken1996,Gardiner2009,VanKampen2007,KondepudiPrigogine2014,SchmiedlSeifert2007,RaoEsposito2016}. Most analytically controlled examples, however, involve a fixed or moderately growing set of species, and often the structure of those species is fixed too, allowing a single arrangement of monomers for a given composition (e.g.\ linear chains). The combinatorial isomeric structure is, in effect, ignored. The two sides have thus been studied apart: one counts the structures that \emph{could} exist, the other follows the kinetics of a \emph{prescribed} few.
	
	Here we bridge them. We consider a setting in which the species are themselves assemblies with hierarchical structure, so the reaction network lives on a combinatorially expanding state space --- not through the monomers, but through the structures in which they arrange. Concretely, we introduce a minimal model in which the combinatorial explosion of possible structures \emph{drives} the dynamics, asking what detailed-balance equilibrium, finite-size fluctuations, and relaxation look like when species are hierarchical assemblies counted by Catalan combinatorics.
	
	The model is intentionally minimal. A closed pool of identical atoms evolves by reversible coagulation and fragmentation \cite{KrapivskyRednerBenNaim2010}; because products form by ordered binary joining, each composite has a unique decomposition and can fragment only into the ordered pair that produced it. Every assembly is therefore a binary tree rather than a linear chain \cite{LahiriEtAl2015}, so assemblies of a given length are counted by the binary parenthesizations of that length, the Catalan numbers \cite{stanley1997enumerative,FlajoletSedgewick2009}. The model is thus simple enough to simulate by Gillespie dynamics \cite{Gillespie1977} yet rich enough to pose a genuinely statistical-mechanical question:  what does detailed-balance equilibrium look like when the number of possible species explodes with size?
	
	The first main point is the deterministic-stochastic organization created by finite volume. At equilibrium the mean copy number of each length-$l$ assembly decays exponentially with $l$, while the number of possible assemblies grows by Catalan combinatorics. Finite volume turns this competition into an observable deterministic-to-stochastic crossover: below a crossover length $l_c$, assemblies are numerous enough that deterministic concentrations are reliable; above $l_c$ individual species are rare and fluctuation-dominated. This is not a trivial finite-size cutoff: because the number of long assemblies grows combinatorially, the rare sector stays statistically important even at large $V$, and its contribution vanishes only slowly with system size. The same structure controls the diversity peak, the longest observed assemblies, and the slow convergence of empirical entropy estimates.
	
	A second point is that the deterministic part of the equilibrium distribution has a Zipf-like rank-frequency form \cite{Zipf1949,Piantadosi2014}. Ranking assemblies by decreasing frequency turns the exponential suppression of long assemblies and the exponential Catalan growth of their number into a power law. In the coagulation-dominated regime the exponent is close to one, a near-Zipf law. This is not produced by a special dynamical mechanism --- preferential attachment, innovation, or optimization \cite{simon1955skew,gabaix1999zipf} --- but follows directly from the combinatorially growing assembly space: exponentially many Catalan assemblies share exponentially small equilibrium probabilities, and the Zipf-like law appears when this neutral hierarchy is viewed in rank order.
	
	The third main point is that the same equilibrium has a thermodynamic interpretation. The system is isothermal, controlled by the single dimensionless parameter $\beta\varepsilon_\mathrm{B}$ built from the bond energy $\varepsilon_\mathrm{B}$ and inverse temperature $\beta=1/T$; one may equivalently fix $T=1$ and vary $\varepsilon_\mathrm{B}$, or fix $\varepsilon_\mathrm{B}=\pm 1$ and vary $T$. Positive $\beta\varepsilon_\mathrm{B}$ favors fragmentation and a monomer-rich pool; negative $\beta\varepsilon_\mathrm{B}$ favors coagulation and large assemblies. At fixed temperature and total atom number each species sees the rest of the pool as a reservoir, giving a grand-canonical description that connects the equilibrium hierarchy to thermodynamic observables --- energy, bond density, entropy. This clarifies the distinction between the thermodynamic entropy of the pool and empirical Shannon entropies from finite samples, whose convergence is strongly affected by the rare sector.
	
	
	Taken together, these points make the model a neutral baseline. No selection rule, catalytic mechanism, or external growth protocol is imposed; all structure comes from detailed balance, mass conservation, and Catalan growth.
	
	
	The paper is organized around these three points. Section~\ref{sec:model} defines the reaction network and its stochastic dynamics. Section~\ref{sec:equilibrium} derives the equilibrium distribution and develops the deterministic-stochastic finite-size crossover and rank-frequency law. Section~\ref{sec:thermodynamics} gives the thermodynamic interpretation. Section~\ref{sec:complexity-measures} collects the entropy, diversity, and assembly-weighted observables that quantify the neutral baseline. Section~\ref{sec:conclusions} closes with perspectives on richer algorithmic-chemistry settings and on relaxation as free expansion in assembly space~\cite{FolenaKruszewski2026FreeExpansion}.

	\section{Model}
	\label{sec:model}
	
	The model describes a closed pool of identical atoms that assemble by ordered binary coagulation and disassemble by the corresponding reverse fragmentation. Since the coagulation is ordered, left and right reactants are distinguished, and each elementary joining can be represented as a branching of a binary tree. Thus each assembly carries a unique recursive binary-tree decomposition. The number of such ordered binary trees with $l$ leaves is the Catalan number $\mathcal{C}_{l-1}$. This combinatorial proliferation of species is the structural fact that drives the non-trivial statistics throughout the paper. We set up the species space, the pool observables, and the reaction network (where uniform mass-action rates make the dynamics neutral) and identify four properties (diluteness, reversibility, loop-freedom, and atom conservation) that together make the model analytically tractable. The deterministic mean-field concentration dynamics that this reaction network generates, and its relaxation to the equilibrium derived here, are the subject of a paper in preparation~\cite{FolenaKruszewski2026FreeExpansion}.

	\subsection{Assemblies}\label{subsec:assemblies}
	
	Assemblies are formed by repeated ordered binary coagulation: the product of $x$ and $y$ is represented by the parenthesized expression $x(y)$ and corresponds to a new ordered binary tree whose left subtree is $x$ and whose right subtree is $y$. The ordering is essential: $x(y)\neq y(x)$ in general. The one-atom assembly is the monomer $\cS$ itself, with length $l(\cS)=1$, where the length $l(x)$ counts the atoms (leaves) of $x$; the number of internal nodes, equal to the number of coagulation events recorded by $x$, is $l(x)-1$.
	
	\begin{figure*}[t]
    \centering
    \begin{tikzpicture}[
        x=1cm,
        y=1cm,
        line cap=round,
        line join=round,
        species/.style={draw=black!45, line width=0.35pt, rounded corners=1pt,
            fill=white, inner xsep=2.4pt, inner ysep=1.8pt, font=\scriptsize},
        active species/.style={draw=black, line width=0.65pt, rounded corners=1pt,
            fill=white, inner xsep=2.4pt, inner ysep=1.8pt, font=\scriptsize},
        product species/.style={draw=black, line width=0.8pt, rounded corners=1pt,
            fill=black!3, inner xsep=3pt, inner ysep=2pt, font=\scriptsize},
        network edge/.style={draw=black!28, line width=0.55pt},
        active edge/.style={draw=black, line width=1.05pt},
        guide/.style={draw=black!18, line width=0.35pt}
    ]
        \node[active species] (S) at (0,0) {$\cS$};
        \node[active species] (A) at (0,-1.35) {$\cS(\cS)$};
        \node[species] (B) at (-1.75,-2.75) {$\cS(\cS(\cS))$};
        \node[active species] (C) at (1.75,-2.75) {$\cS(\cS)(\cS)$};
        \node[species] (D) at (-4.25,-4.15) {$\cS(\cS(\cS(\cS)))$};
        \node[species] (E) at (-2.15,-4.15) {$\cS(\cS(\cS)(\cS))$};
        \node[species] (F) at (0,-4.15) {$\cS(\cS)(\cS(\cS))$};
        \node[species] (G) at (2.15,-4.15) {$\cS(\cS(\cS))(\cS)$};
        \node[species] (H) at (4.25,-4.15) {$\cS(\cS)(\cS)(\cS)$};
        \node[product species] (T) at (1.75,-5.75) {$\cS(\cS)(\cS(\cS)(\cS))$};

        \node[font=\scriptsize] at (-5.35,0) {$l=1$};
        \node[font=\scriptsize] at (-5.35,-1.35) {$l=2$};
        \node[font=\scriptsize] at (-5.35,-2.75) {$l=3$};
        \node[font=\scriptsize] at (-5.35,-4.15) {$l=4$};
        \node[font=\scriptsize] at (-5.35,-5.75) {$l=5$};
        \draw[guide] (-4.9,-0.55) -- (4.9,-0.55);
        \draw[guide] (-4.9,-2.05) -- (4.9,-2.05);
        \draw[guide] (-4.9,-3.45) -- (4.9,-3.45);
        \draw[guide] (-4.9,-4.95) -- (4.9,-4.95);

        \draw[network edge] (S) -- (A);
        \draw[network edge] (S) -- (B);
        \draw[network edge] (A) -- (B);
        \draw[network edge] (A) -- (C);
        \draw[network edge] (S) -- (C);
        \draw[network edge] (S) -- (D);
        \draw[network edge] (B) -- (D);
        \draw[network edge] (S) -- (E);
        \draw[network edge] (C) -- (E);
        \draw[network edge] (A) -- (F);
        \draw[network edge] (B) -- (G);
        \draw[network edge] (S) -- (G);
        \draw[network edge] (C) -- (H);
        \draw[network edge] (S) -- (H);

        \draw[active edge] (S) -- (A);
        \draw[active edge] (A) -- (C);
        \draw[active edge] (S) -- (C);
        \draw[active edge] (A) -- (T);
        \draw[active edge] (C) -- (T);

        \begin{scope}[xshift=7.0cm,yshift=-0.25cm]
            \node[font=\scriptsize] at (0,0.35) {assembly: binary tree};
            \coordinate (r) at (0,-0.25);
            \coordinate (l) at (-1.25,-1.25);
            \coordinate (rr) at (1.25,-1.25);
            \coordinate (ll) at (-1.85,-2.25);
            \coordinate (lr) at (-0.65,-2.25);
            \coordinate (rl) at (0.65,-2.25);
            \coordinate (rs) at (1.85,-2.25);
            \coordinate (rll) at (0.1,-3.25);
            \coordinate (rlr) at (1.2,-3.25);
            \draw[active edge] (r) -- (l);
            \draw[active edge] (r) -- (rr);
            \draw[active edge] (l) -- (ll);
            \draw[active edge] (l) -- (lr);
            \draw[active edge] (rr) -- (rl);
            \draw[active edge] (rr) -- (rs);
            \draw[active edge] (rl) -- (rll);
            \draw[active edge] (rl) -- (rlr);
            \node[product species] at (r) {$\cS(\cS)(\cS(\cS)(\cS))$};
            \node[active species] at (l) {$\cS(\cS)$};
            \node[active species] at (rr) {$\cS(\cS)(\cS)$};
            \node[active species] at (rl) {$\cS(\cS)$};
            \node[active species] at (ll) {$\cS$};
            \node[active species] at (lr) {$\cS$};
            \node[active species] at (rll) {$\cS$};
            \node[active species] at (rlr) {$\cS$};
            \node[active species] at (rs) {$\cS$};
        \end{scope}
    \end{tikzpicture}
    \caption{Layered embedding of the reaction network of assemblies up to length $l=4$. The highlighted subnetwork (black edges) leads to a length-$5$ assembly, written as the parenthesized expression $\cS(\cS)(\cS(\cS)(\cS))$ and represented on the right as a binary tree. Light gray edges show the other ordered compositions available in the same local region of the combinatorial space.}
    \label{fig:catalan-assembly-network}
\end{figure*}
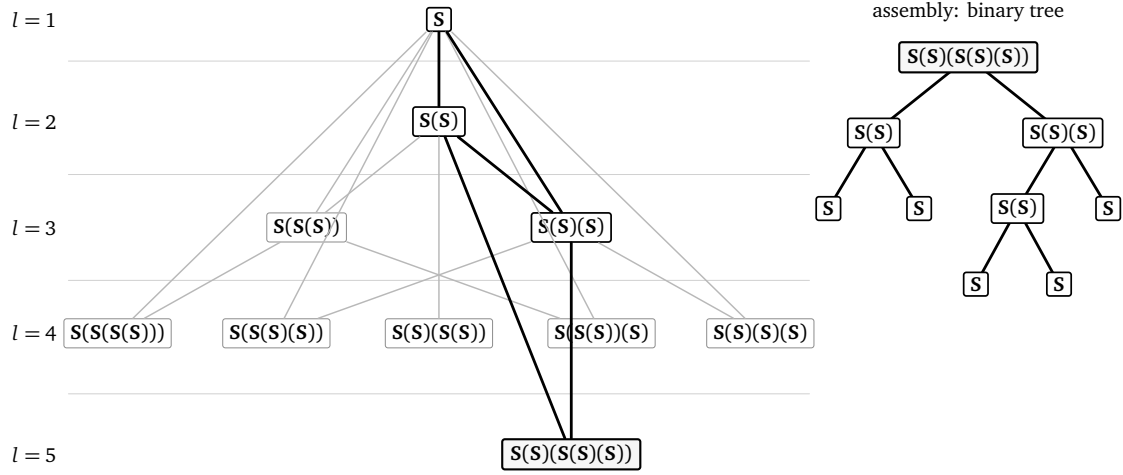

	The number of distinct assemblies of length $l$ is given by the Catalan number
	\begin{equation}
		\mathcal{N}_l = \mathcal{C}_{l-1} \ ,
		\qquad
		\mathcal{C}_k = \frac{1}{k+1}\binom{2k}{k} \ .
		\label{eq:catalan-count}
	\end{equation}
	The first ten values are
	\[
	\mathcal{N}_l = 1,\,1,\,2,\,5,\,14,\,42,\,132,\,429,\,1430,\,4862 \ ,
	\]
	with the large-$l$ asymptotics
	\begin{equation}
		\mathcal{N}_l \sim \frac{4^{\,l-1}}{\sqrt{\pi}\,(l-1)^{3/2}} \ .
	\end{equation}
	
	Figure~\ref{fig:catalan-assembly-network} shows the portion of the assembly network generated by coagulation events up to length $l=4$, starting from the atomic species $S$. Figure~\ref{fig:binary-tree-orders} lists all binary-tree assemblies up to $l=5$, grouped by left-right symmetry class. Equivalent planar-tree and Dyck-path representations of the parenthesized expression are given in Appendix~\ref{app:assembly-representations}.

	\subsection{Pool}\label{subsec:pool}
	
	We consider a closed pool with a fixed number of constituent atoms $\cS$. The atom density is set to one, so a volume $V$ contains total atomic mass $V$\footnote{This choice fixes the unit of volume so that there is on average one atom per unit volume; any other atom concentration can be rescaled to this convention.}. Since every allowed reaction conserves atoms, the total atom count is
	\begin{equation}
		M\equiv\sum_x l(x)\, n_x = V \ .
		\label{eq:mass-conservation}
	\end{equation}
	The pool is then specified by its occupation-number vector,
	\begin{equation}
		\bm{n}=(n_{x_1},n_{x_2},\dots,n_{x_\mathcal{D}}), \qquad \sum_x n_x = N \ ,
	\end{equation}
	where $\mathcal{D}$ is the diversity of the system, namely the number of distinct assemblies present in the pool, and $N$ is the total number of assemblies. Equivalently, we may denote the microscopic state by $\bm{n}=\{n_x\}$, with $n_x=0$ for absent assemblies; this notation is analytically convenient even though most entries are zero, since the species space grows combinatorially. The concentration of assembly $x$ takes two forms: the empirical concentration $n_x/V$ and the expected concentration 
	\begin{equation}
		c_x = \frac{\langle n_x \rangle}{V} \ ,
	\end{equation}where $\langle  \rangle$ in this manuscript will refer to the expectation at equilibrium.
	For non-rare species, namely those with $c_x V \gg 1$, because of central limit theorem, the two almost coincide ; however $c_x V \sim 1$ this is not the case; the breakdown of this self-averaging for rare species is the central object of Sec.~\ref{sec:equilibrium}.
	
	\begingroup
\setlength{\tabcolsep}{2pt}
\setlength{\arrayrulewidth}{0.25pt}
\renewcommand{\arraystretch}{1.02}
\newcommand{\treeentry}[3][0.67]{%
    \scalebox{#1}{%
        \begin{tabular}[t]{@{}c@{}}
            \bttree{#2} \\[-2pt]
            $\scriptstyle #3$
        \end{tabular}%
    }%
}
\newcommand{\treeatom}[2][0.67]{%
    \scalebox{#1}{%
        \begin{tabular}[t]{@{}c@{}}
            \btdot \\[-2pt]
            $\scriptstyle #2$
        \end{tabular}%
    }%
}
\begin{figure*}[t]
    \centering
    \resizebox{\textwidth}{!}{%
    \begin{tabular}{|c|c|}
        \hline
        \textbf{Length} & \textbf{Assemblies -- binary-trees and expressions} \\
        \hline
        \textbf{1}
        &
        \treeatom{\cS}
        \\
        \hline
        \textbf{2}
        &
        \treeentry{[[][]]}{\cS(\cS)}
        \\
        \hline
        \textbf{3}
        &
        \begin{tabular}[t]{@{}c@{\hspace{0.45em}}c@{}}
            \treeentry{[[][[][]]]}{\cS(\cS(\cS))} &
            \treeentry{[[[][]][]]}{\cS(\cS)(\cS)}
        \end{tabular}
        \\
        \hline
        \textbf{4}
        &
        \begin{tabular}[t]{@{}cc|c|cc@{}}
            \treeentry{[[][[][[][]]]]}{\cS(\cS(\cS(\cS)))} &
            \treeentry{[[][[[][]][]]]}{\cS(\cS(\cS)(\cS))} &
            \treeentry{[[[][]][[][]]]}{\cS(\cS)(\cS(\cS))} &
            \treeentry{[[[][[][]]][]]}{\cS(\cS(\cS))(\cS)} &
            \treeentry{[[[[][]][]][]]}{\cS(\cS)(\cS)(\cS)}
        \end{tabular}
        \\
        \hline
        \textbf{5}
        &
        \begin{tabular}[t]{@{}cccc|cc|cc|cc|cccc@{}}
            \treeentry{[[][[][[][[][]]]]]}{\cS(\cS(\cS(\cS(\cS))))} &
            \treeentry{[[][[][[[][]][]]]]}{\cS(\cS(\cS(\cS)(\cS)))} &
            \treeentry{[[][[[][[][]]][]]]}{\cS(\cS(\cS(\cS))(\cS))} &
            \treeentry{[[][[[[][]][]][]]]}{\cS(\cS(\cS)(\cS)(\cS))} &
            \treeentry{[[[][]][[][[][]]]]}{\cS(\cS)(\cS(\cS(\cS)))} &
            \treeentry{[[[][]][[[][]][]]]}{\cS(\cS)(\cS(\cS)(\cS))} &
            \treeentry{[[][[[][]][[][]]]]}{\cS(\cS(\cS)(\cS(\cS)))} &
            \treeentry{[[[[][]][[][]]][]]}{\cS(\cS)(\cS(\cS))(\cS)} &
            \treeentry{[[[][[][]]][[][]]]}{\cS(\cS(\cS))(\cS(\cS))} &
            \treeentry{[[[[][]][]][[][]]]}{\cS(\cS)(\cS)(\cS(\cS))} &
            \treeentry{[[[][[][[][]]]][]]}{\cS(\cS(\cS(\cS)))(\cS)} &
            \treeentry{[[[][[[][]][]]][]]}{\cS(\cS(\cS)(\cS))(\cS)} &
            \treeentry{[[[[][[][]]][]][]]}{\cS(\cS(\cS))(\cS)(\cS)} &
            \treeentry{[[[[[][]][]][]][]]}{\cS(\cS)(\cS)(\cS)(\cS)}
        \end{tabular}
        \\
        \hline
    \end{tabular}
    }
    \caption{Binary-tree assemblies up to length $l=5$. The ordered trees are arranged symmetrically, and thin vertical separators indicate left-right symmetry classes. Each tree is labeled by its parenthesized expression.}
    \label{fig:binary-tree-orders}
\end{figure*}
\endgroup

	Given the concentration $c_x$, the probability of drawing assembly $x$ uniformly from the pool of $N$ assemblies is
	\begin{equation}
		p_x = \rho^{-1} c_x \qquad \text{with} \quad\rho=N/V,
	\end{equation}
	where $\rho$ is the total assembly density. Since the initial atom density is $1$ and non-monomer assemblies contain more than one atom, $\rho<1$ whenever such assemblies are present.

	\subsection{Reaction Network}
	
	The pool evolves by a reversible dilute coagulation/fragmentation reaction,
	\begin{equation}
		x + y
		\xrightleftharpoons[k_F]{k_K}
		x(y),
		\label{eq:neutral-reaction}
	\end{equation}
	where $x(y)$ is the ordered binary assembly obtained by joining assemblies $x$ and $y$. The associated stochastic dynamics is defined by the mass-action transition probabilities \cite{Gardiner2009}
	\begin{equation}
		w_{x+y\to x(y)} = k_K \frac{n_x n_y}{V},
		\qquad
		w_{x(y)\to x+y} = k_F\, n_{x(y)} ,
		\label{eq:Gillespie}
	\end{equation}
	in which the coagulation and fragmentation rates $k_K$ and $k_F$ are independent of the specific reacting assemblies, so the dynamics has no bias toward any particular assembly. At finite $V$, when the two reactants are identical, the ordered-pair count $n_x(n_x-1)$ replaces $n_x^2$ to avoid self-pairing. The Gillespie algorithm used to simulate this dynamics is described in Appendix~\ref{app:gillespie-sim}.
	
	Two structural features distinguish this reaction from generic binary coagulation and fragmentation. First, it is \textit{ordering-dependent}: the two reactants enter the product asymmetrically as left and right subtrees. Second, it is \textit{encapsulating}: each composite is built by wrapping one assembly, in the parenthesized notation, as the right argument of another, so the only immediate fragmentation available to $x(y)$ is the one that opens its outermost parenthesis, recovering $x$ and $y$ (cf.\ Fig.~\ref{fig:catalan-assembly-network}). The binary-tree representation of an assembly therefore doubles as the reaction graph that builds it from atoms (each leaf is an atom $\cS$), as illustrated in Fig.~\ref{fig:binary-tree-orders}.
	
	Together with the underlying coagulation/fragmentation form, these features yield four structural properties that make the model analytically tractable:
	\begin{enumerate}
		\item \textit{dilute}: the system is well mixed, with no explicit spatial structure, so interactions depend only on concentrations;
		\item \textit{reversible}: every reaction can proceed in both directions;
		\item \textit{loop-free}: each assembly has a unique immediate decomposition (a direct consequence of the encapsulating structure), so the reaction graph has no alternative channel producing the same composite assembly;
		\item \textit{atom-conserving}: every reaction preserves the total number of atoms.
	\end{enumerate}
	Together, these four properties yield an essentially complete analytical characterization of equilibrium, developed in Sec.~\ref{sec:equilibrium}.

	\begin{table}[t]
		\centering
		\small
		\begin{tabular}{lll}
			\hline
			Symbol & Meaning & First introduced \\
			\hline
			$l(x)$ & length of assembly $x$ & Sec.~\ref{subsec:assemblies} \\
			$V$ & system volume, equal to the initial atom count & Sec.~\ref{subsec:pool} \\
			$M=\sum_x l(x)n_x=V$ & conserved total mass & Sec.~\ref{subsec:pool} \\
			$n_x$, $c_x$ & occupation number and concentration of assembly $x$ & Sec.~\ref{subsec:pool} \\
			$N=\sum_x n_x$, $\rho=\sum_x c_x$ & total number and density of assemblies & Sec.~\ref{subsec:pool} \\
			$p_x=c_x/\rho$ & relative frequency of assembly $x$ & Sec.~\ref{subsec:pool} \\
			$\mathcal{D}$ & diversity (number of distinct assemblies present) & Sec.~\ref{subsec:pool} \\
			$r=k_F/k_K$, $q=c_\cS/r$ & kinetic ratio and equilibrium fugacity & Sec.~\ref{subsec:generating-function} \\
			$B=M-N=V-N$ & total number of internal bonds in the closed pool & Sec.~\ref{sec:thermodynamics} \\
			$\varepsilon_\mathrm{B},\mu_\cS$ & bond energy and atomic chemical potential & Sec.~\ref{subsec:bond-chemical} \\
			$S^{\mathrm{th}}$ & thermodynamic entropy & Sec.~\ref{sec:thermodynamics} \\
			$s^{\mathrm{Sh}}$, $S^{\mathrm{Sh}}_{\mathrm{pool}}$ & single-assembly and pool Shannon entropies & Sec.~\ref{subsec:entropy} \\
			$\mathcal{A}$ & assembly-weighted observables & Sec.~\ref{sec:complexity-measures} \\
			\hline
		\end{tabular}
		\caption{Main notation used throughout the paper.}
		\label{tab:notation}
	\end{table}
	
	\section{Equilibrium and Finite-Size Crossover}
	\label{sec:equilibrium}
	
	We now describe the equilibrium organization of the assembly pool, focusing on the structural picture that emerges from the equilibrium distribution and finite-size combinatorics. The chemical-network thermodynamics built on the same equilibrium is developed in Sec.~\ref{sec:thermodynamics}.
	
	The starting point is the exact equilibrium distribution. Local detailed balance on a loop-free reaction network, combined with mass conservation over the combinatorially growing Catalan species space, fixes the concentration of an assembly of length $l$ to the simple form $c_l = r q^l$, where $r = k_F/k_K$ is the kinetic ratio and $q(r)$ the equilibrium fugacity (Secs.~\ref{subsec:equilibrium-solution}--\ref{subsec:generating-function}).
	
	The rest of the section is organized around the competition between the exponential decay of each single-species concentration, $c_l \propto q^l$, and the Catalan growth of the number of species, $\mathcal{N}_l \sim 4^l/l^{3/2}$. This competition defines a finite-size crossover length $l_c \sim \ln V$ between a deterministic high-copy sector and a stochastic rare-species sector, together with a family of related length scales. In the deterministic sector, the same balance produces a Zipf-like rank-frequency law. Beyond the crossover, it explains the persistent macroscopic stochasticity of the coagulation-dominated regime ($r<1$), as well as the subextensive growth of diversity and assembly-weighted observables developed in Sec.~\ref{sec:complexity-measures}.

	\subsection{Equilibrium Solution}\label{subsec:equilibrium-solution}
	To derive the equilibrium distribution $\mathbb{P}^{\mathrm{eq}}(\bm{n})$, we impose local detailed balance on each elementary reaction channel $x + y\rightleftharpoons x(y)$,
	\begin{equation}
		\mathbb{P}^{\mathrm{eq}}(\bm{n}+\bm{1}_x+\bm{1}_y)\, w_{x+y\to x(y)} = \mathbb{P}^{\mathrm{eq}}(\bm{n}+\bm{1}_{x(y)})\, w_{x(y)\to x+y} ,
		\label{eq:detailed-balance}
	\end{equation}
	where $\bm{1}_x$ is the occupation vector with a $1$ at position $x$ and zeros elsewhere, and each propensity is evaluated at the corresponding state. This condition states that the equilibrium probability flux along the forward direction of every reaction is exactly compensated by that of its reverse. In the grand-canonical representation, where the total mass is fixed in expectation, the equilibrium probability $\mathbb{P}^{\mathrm{eq}}(\bm{n})$ factorizes over assembly types $x$, with factorial weights $n_x!$ for indistinguishable copies. The result is a multivariate Poisson distribution \cite{SchmiedlSeifert2007},
	\begin{equation}\label{eq:equilib_dist}
		\mathbb{P}^{\mathrm{eq}}(\bm{n}) = \prod_x e^{-\langle n_x\rangle}\,\frac{\langle n_x\rangle^{n_x}}{n_x!} ,
	\end{equation}
	where $\langle n_x\rangle = \langle N \rangle\, p^{\mathrm{eq}}_x = V c^{\mathrm{eq}}_x$ is the expected number of assemblies of type $x$, $\langle N \rangle$ is the expected total number of assemblies, and $V$ is the imposed mean total atomic mass \footnote{For a strictly closed finite-volume pool, the conservation law in Eq.~\eqref{eq:mass-conservation} must be imposed exactly; the equilibrium measure is then the same product-Poisson law conditioned on $\sum_x l(x)n_x=V$. This conditioning only changes the normalization inside the conserved mass shell and therefore leaves the local detailed-balance ratios unchanged ~\cite[Sec.~11.5.1]{Gardiner2009}. }. Substituting the Poisson form into Eq.~\eqref{eq:detailed-balance} yields the concentration relation
	\begin{equation}
		\frac{c_{x(y)}}{c_x c_y} = \frac{k_K}{k_F}, \qquad \forall \quad x + y\rightleftharpoons x(y),
		\label{eq:local-db}
	\end{equation}
	where the superscript $\mathrm{eq}$ is omitted whenever equilibrium is understood from context. Equivalently, in terms of pool probabilities $p_x=c_x/\rho$, this reads $p_{x(y)}/(p_xp_y)=\rho k_K/k_F$. Because the reaction network is loop-free, Eq.~\eqref{eq:local-db} determines the entire concentration hierarchy once the monomer concentration $c_\cS$ is fixed by the global mass constraint. For example, $c_{\cS(\cS)} = (k_K/k_F)c_\cS^2$ and $c_{\cS(\cS)(\cS)} = (k_K/k_F)^2c_\cS^3 = c_{\cS(\cS(\cS))}$, illustrating that the result depends only on length, not on tree shape. More generally,
	\begin{equation}\label{eq:equil-p_l}
		c_l = \bigl(k_K/k_F\bigr)^{l-1}\,c_\cS^l ,
	\end{equation}
	where $c_l$ denotes the concentration of any assembly of length $l$ (so $c_1=c_\cS$). The corresponding single-assembly probability is $p_l=c_l/\rho$. The actual value of $c_\cS$ requires summing this hierarchy against the mass conservation Eq.~\eqref{eq:mass-conservation} on the combinatorially growing network, which we carry out in the next subsection.

	\subsection{Generating Function}\label{subsec:generating-function}
	
	We now compute the equilibrium concentration $c_\cS$ by imposing mass conservation to the hierarchy of concentrations in Eq.~\ref{eq:equil-p_l}. For simplicity we define the kinetic ratio
	\begin{equation}
		r=\frac{k_F}{k_K} \qquad\text{and the fugacity}\qquad q=\frac{c_\cS}{r} \ .
	\end{equation}
	Thus the concentration of one assembly of length $l$ takes the simple form
	\begin{equation}
		c_l = r\, q^l .
	\end{equation}
	The fugacity $q$ is fixed by the mass-conservation condition \ref{eq:mass-conservation},
	\begin{equation}
		1 = \sum_x l(x)\, c_x = \sum_{l\geq 1} \mathcal{N}_l\, l\, c_l
		= r \sum_{l\geq 1}  \mathcal{N}_{l}\, l\, q^l .
	\end{equation}
	The second equality uses the fact that at equilibrium the concentration of an assembly $x$ depends only on its length $l(x)$ and that, for each length $l$, there are $\mathcal{N}_{l}$ assembly types. It is then convenient to introduce the generating function of the assembly space,
	\begin{equation}\label{eq:gen-func}
		G_\mathcal{N}(u)=\sum_{l\geq 1} \mathcal{N}_{l}\, u^l=\frac{1-\sqrt{1-4u}}{2}.
	\end{equation}
	The series converges on $0<u<1/4$, where $G_\mathcal{N}(u)$ increases monotonically from $0$ to $1/2$ at the singular point $u=1/4$; this object encodes the combinatorial growth of the assembly space\footnote{Equivalently, the generating function of the Catalan numbers themselves is $G_\mathcal{C}(u)=\sum_{k\geq 0}\mathcal{C}_{k}\,u^k = G_\mathcal{N}(u)/u$.}. Using $\sum_{l\geq 1} l\, \mathcal{N}_{l}\, q^l = q\,G_\mathcal{N}'(q)$, the mass-conservation condition becomes
	\begin{equation}\label{eq:consV}
		1 = r q\, G_\mathcal{N}'(q) = \frac{r q}{\sqrt{1-4q}},
	\end{equation}
	which fixes
	\begin{equation}
		q(r) = \frac{1}{2+\sqrt{4+r^2}} .
	\end{equation}
	In what follows the $r$-dependence of $q$ is kept implicit. The total assembly density and the mean length follow immediately,
	\begin{equation}
		\rho = \sum_{l\geq 1} \mathcal{N}_l\, c_l = r\, G_\mathcal{N}(q) ,
		\qquad
		\overline{l} \equiv \sum_{l\geq 1}\mathcal{N}_l\, l\, p_l = \frac{1}{\rho},
		\label{eq:avg-length}
	\end{equation}
	where $p_l = c_l/\rho$ is the probability of one assembly of length $l$ and the overline denotes the equilibrium average over the assemblies in the pool.
	
	As a reference point, $r=1$ gives $q = 1/(2+\sqrt{5}) \approx 0.236$ and therefore $c_l \approx 0.236^l$, in agreement with the equilibrium Gillespie snapshots analyzed below. Unless explicitly stated otherwise, all plots below are evaluated at this point:\begin{equation}
		r=1, \qquad q = \frac{1}{2+\sqrt{5}}.
	\end{equation}

	\begin{figure}[t]
		\centering
		\includegraphics[width=\columnwidth]{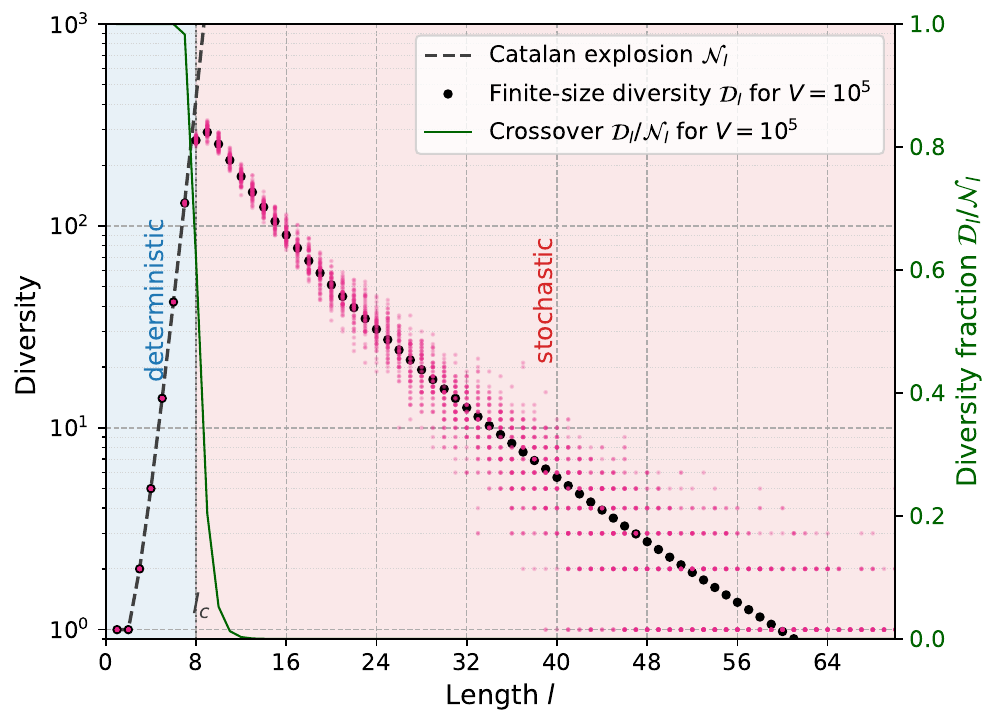}
		\caption{Finite-size diversity by length for $V=10^5$ at $r=1$. Black points show the expected diversity $\langle\mathcal{D}_l\rangle$, the dashed gray curve gives the Catalan count $\mathcal{N}_l=\mathcal{C}_{l-1}$, and the dark-green curve shows the occupancy ratio $\mathcal{D}_l/\mathcal{N}_l$ across the deterministic-to-stochastic crossover at $l_c$. Purple dots show the diversity of 60 simulation samples for $V=10^5$. The diversity peak lies close to the crossover scale $l_c$.}
		\label{fig:finite-size-diversity}
	\end{figure}

	\subsection{Finite-size Crossover}
	The equilibrium pool contains exponentially fewer copies of each individual assembly as the length grows, since
	\[
	\langle n_l\rangle = V c_l = V r q^l .
	\]
	At the same time, the number of possible assemblies of length $l$ grows combinatorially,
	\[
	\mathcal{N}_l \sim \frac{4^{\,l-1}}{\sqrt{\pi}\,(l-1)^{3/2}} .
	\]
	The competition between these two trends defines a natural crossover length at which the mean occupancy per species drops to unity,
	\begin{equation}
		\langle n_{l_c}\rangle = 1 ,
	\end{equation}
	giving
	\begin{equation}
		l_c = -\frac{\ln(Vr)}{\ln q}
		= \frac{\ln(Vr)}{\ln\!\left(2+\sqrt{4+r^2}\right)} .
		\label{eq:lc}
	\end{equation}
	This logarithmically growing scale separates a deterministic sector ($l\ll l_c$), where typical occupancies are large, from a stochastic sector ($l\gg l_c$), where individual assemblies appear only in a few copies.
	
	The crossover is directly visible in the length-resolved diversity. Since each occupancy is Poisson-distributed, the probability that a given assembly of length $l$ is present at least once is $1-e^{-\langle n_l\rangle}$. The expected length-resolved diversity, the number of distinct assemblies of length $l$ realized in the pool, is therefore
	\begin{equation}
		\langle\mathcal{D}_l\rangle
		= \mathcal{N}_l\bigl(1-e^{-\langle n_l \rangle}\bigr) .
		\label{eq:diversity-single}
	\end{equation}
	For $l\ll l_c$ one has $\langle n_l\rangle\gg1$, almost every allowed assembly is present, and $\mathcal{D}_l\simeq \mathcal{N}_l$. For $l\gg l_c$ one has $\langle n_l\rangle\ll1$, only a small fraction is realized, and $\mathcal{D}_l\simeq \mathcal{N}_l\,\langle n_l\rangle$. Figure~\ref{fig:finite-size-diversity} illustrates this at $V=10^5$ and $r=1$: the diversity peaks at $l_*=9$, close to $l_c \approx 7.98$, with the visible fraction collapsing rapidly past the crossover. The diversity is itself a stochastic quantity with relative variance
	\begin{equation}
		\frac{\text{Var}(\mathcal{D}_l)}{\langle\mathcal{D}_l\rangle^2}
		= \frac{e^{-\langle n_l \rangle}}{\langle\mathcal{D}_l\rangle}
		= \frac{1}{\mathcal{N}_l\bigl(e^{\langle n_l \rangle}-1\bigr)} .
	\end{equation}
	Below $l_c$ these relative fluctuations are exponentially small in $V$; above $l_c$ they scale as $1/\langle\mathcal{D}_l\rangle$. Both regimes are visible in Fig.~\ref{fig:finite-size-diversity} (purple points from 60 samples).
	
	\begin{figure}[t]
		\centering
		\includegraphics[width=\columnwidth]{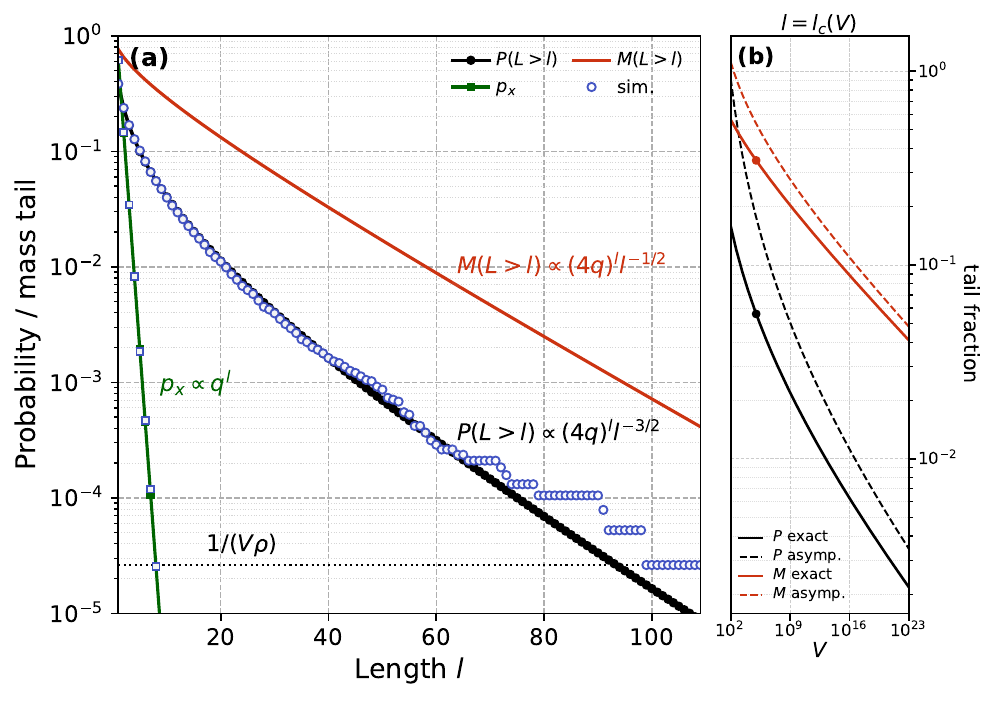}
		\caption{Tail probability, single-assembly probability, and tail mass at $r=1$. \textbf{(a)} Tail probability $P(L\!>\!l)$ (black), single-assembly probability $p_l$ (dark green), and tail mass $M(L\!>\!l)$ (red dashed) plotted against the assembly length $l$. Blue open  markers show a Gillespie snapshot $V=10^5$. The horizontal dotted line marks the scale $1/(V\rho)=1/N$, and the labels indicate the corresponding large-$l$ asymptotic forms. \textbf{(b)} Exact and asymptotic scaling of $P(L\!>\!l_c)$ and $M(L\!>\!l_c)$ with the volume $V$. The dot indicates the value for $V=10^5$.}
		\label{fig:tail-probability}
	\end{figure}
	
	\subsection{Macroscopic Stochasticity}\label{subsec:macroscopic_stochasticity}
	
	The crossover $l_c$ is defined at the level of a single species. To see whether the rare sector forms only a microscopic tail or a macroscopic part of the pool, we ask what fraction of assemblies and of atomic mass lies beyond $l_c$. Let $L$ denote the length of an assembly drawn at random from the pool. We define the tail probability
	\begin{equation}
		P(L>l)=\sum_{n\ge l+1}\mathcal{N}_n\, p_n ,
	\end{equation}
	which counts the fraction of assemblies longer than $l$, and the tail mass
	\begin{equation}
		M(L>l)=\sum_{n\ge l+1} n\,\mathcal{N}_n\, c_n ,
	\end{equation}
	which measures the fraction of total atomic mass carried by those long assemblies. Both are shown at $r=1$ in Fig.~\ref{fig:tail-probability}(a), together with finite-size simulations at $V=10^5$. For large $l$, both sums are dominated by their first few terms ($n\approx l$), so the mass tail acquires an extra factor $\sim l\rho$ relative to the probability tail. Using the Catalan asymptotic in the leading-order geometric sum,
	\begin{equation}
		P(L>l)\sim \frac{M(L>l)}{l\rho} \sim \frac{rq}{\rho\sqrt{\pi}(1-4q)}\frac{(4q)^l}{l^{3/2}} .
	\end{equation}
	Defining
	\begin{equation}
		\alpha = -\frac{\ln q}{\ln 4} > 1 ,
		\label{eq:alpha-def}
	\end{equation}
	which at the reference point $r=1$ gives
	\begin{equation}
		\alpha \approx 1.041 ,
		\label{eq:alpha-ref}
	\end{equation}
	and using $(4q)^{l_c} = (Vr)^{1/\alpha-1}$, evaluation at $l=l_c(V)$ gives the fraction of the system that lies in the stochastic regime,
	\begin{equation}
		P(L>l_c) \sim \frac{M(L>l_c)}{l_c\,\rho}
		\sim
		\frac{(Vr)^{1/\alpha-1}}{[\ln(Vr)]^{3/2}} .
		\label{eq:tail-fraction-lc}
	\end{equation}
	Both fractions vanish as $V\to\infty$, but the convergence is slow in the coagulation-dominated regime $q\to 1/4$, where $\alpha\to 1$ and the decay becomes logarithmic. In particular, Fig.~\ref{fig:tail-probability}(b) shows that at $r=1$, even at volume $V=10^{23}$ about $4\%$ of the atomic mass still lies above $l_c$. A non-negligible fluctuating sector therefore survives at macroscopic volumes.
	
	\begin{figure}[t]
		\centering
		\includegraphics[width=\columnwidth]{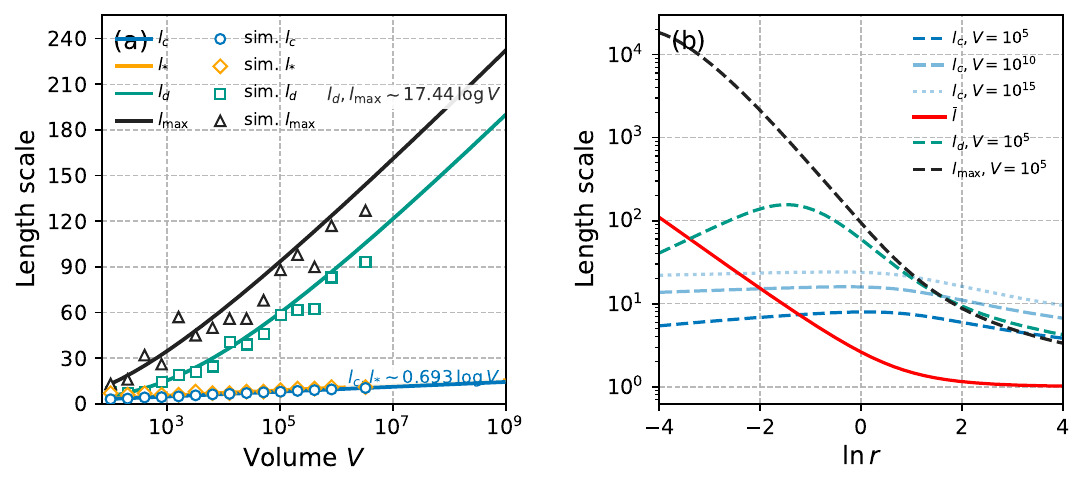}
		\caption{Characteristic finite-size lengths. \textbf{(a)} Volume scaling at $r=1$. Solid curves show the occupancy crossover $l_c$, the diversity peak $l_*$, the diversity cutoff $l_d$, and the largest observed length $l_{\max}$. Open symbols show measurements extracted from Gillespie snapshots at time $t_f=50$ for different volumes $V$; the empirical $l_d,l_{\max}$ are systematically underestimated if $t_f$ is not long enough. The pair $l_c,l_*$ grows with slope $-1/\ln q$, while $l_d$ and $l_{\max}$ share the larger slope $-1/\ln(4q)$. \textbf{(b)} Dependence on $\ln r$ at $\beta=1$, compared with the expected equilibrium mean length $\overline{l}$ (volume independent) . The dashed curves use the same color code as in \textbf{(a)}, with $l_c$ shown for $V=10^5,10^{10},10^{15}$ and $l_d,l_{\max}$ shown for $V=10^5$.}
		\label{fig:length-scales}
	\end{figure}
	
	\subsection{Finite-size Lengths}
	\label{sec:finite-size-lengths}
	The same crossover picture naturally introduces four characteristic lengths. The first is the occupancy scale $l_c$ of Eq.~\eqref{eq:lc}. The second is the position $l_*$ of the peak of $\mathcal{D}_l$. As shown in Appendix~\ref{app:length-scale-asymptotics},
	\begin{equation}
		l_* \sim l_c \sim -\frac{\ln(Vr)}{\ln q}
		\approx 0.693\,\ln V \quad (\text{at } r=1) ,
	\end{equation}
	so the diversity peak follows the same leading logarithmic drift as the occupancy crossover. The third scale is the diversity cutoff $l_d$, defined by
	\begin{equation}
		\langle \mathcal{D}_{l_d} \rangle = 1 ,
		\label{eq:ld-condition}
	\end{equation}
	namely the length at which the descending branch of the average diversity profile reaches order one. The fourth is the largest observed length $l_{\max}$ in a pool of mean size $N=V\rho$, defined by
	\begin{equation}
		V\rho\, P(L>l_{\max}) = 1 ,
		\label{eq:lmax-condition}
	\end{equation}
	namely the length beyond which the expected number of assemblies in the tail becomes of order one. Since $4q<1$, the two tail-controlled scales share a common leading behavior,
	\begin{equation}
		l_d \sim l_{\max} \sim -\frac{\ln V}{\ln(4q)}
		\approx 17.44\,\ln V \quad (\text{at } r=1) .
	\end{equation}
	Thus $l_c$ and $l_*$ are set by the single-species decay $-\ln q$, whereas $l_d$ and $l_{\max}$ are set by the steeper tail scale $-\ln(4q)$, growing markedly faster, by a factor of about $25$ at $r=1$. Figure~\ref{fig:length-scales}(a) compares these scales with Gillespie estimates, with the asymptotic slopes indicated on the plot; the full asymptotic expansions are collected in Appendix~\ref{app:length-scale-asymptotics}. All four lengths grow without bound as $V\to\infty$, in contrast to the mean length $\overline{l}=1/\rho$ of Eq.~\eqref{eq:avg-length}, which is a thermodynamic observable and remains finite in the infinite-volume limit.
	
	Figure~\ref{fig:length-scales}(b) compares these finite-size scales with $\overline{l}$ as the coagulation/fragmentation balance is varied. Depending on where $\overline{l}$ sits relative to $l_c$, $l_d$, $l_{\max}$, three qualitatively distinct stochastic regimes appear:
	\begin{itemize}
		\item \textit{Tail-stochastic} ($\overline{l}\ll l_c$, weak coagulation $r\gg 1$): stochasticity is confined to the long-assembly tail; typical assemblies are deterministic.
		\item \textit{Bulk-stochastic} ($\overline{l}\gtrsim l_c$, stronger coagulation): typical assemblies lie in the rare-copy sector, so the bulk itself is fluctuation-dominated and no longer well described by deterministic concentrations.
		\item \textit{Small-number stochastic} ($\overline{l}\sim l_d, l_{\max}$, very strong coagulation): only a handful of assemblies occupy the relevant length sector. The limiting case is $\langle N\rangle=V\rho=O(1)$, equivalently $\overline{l}=O(V)$, where a single assembly can carry an $O(1)$ fraction of all atoms. In Sec. \ref{subsec:thermal-bath} we will see that this is a condensation crossover in temperature $T_c\sim1/\ln V$.
	\end{itemize}
	
	\subsection{Rank-frequency and Zipf-law}
	The equilibrium hierarchy can also be read as a rank-frequency law. Ordering assemblies by decreasing probability gives the standard representation familiar, for example, from word frequencies in a text. In many systems such plots are approximately power laws, often with an exponent close to one; this is the classical Zipf law. Here the Zipf-like behavior is not imposed by a growth rule: it emerges because exponentially many Catalan species share exponentially small equilibrium probabilities, while finite volume truncates the law at the crossover where individual copy numbers become rare.

	\begin{figure}[t]
		\centering
		\includegraphics[width=\columnwidth]{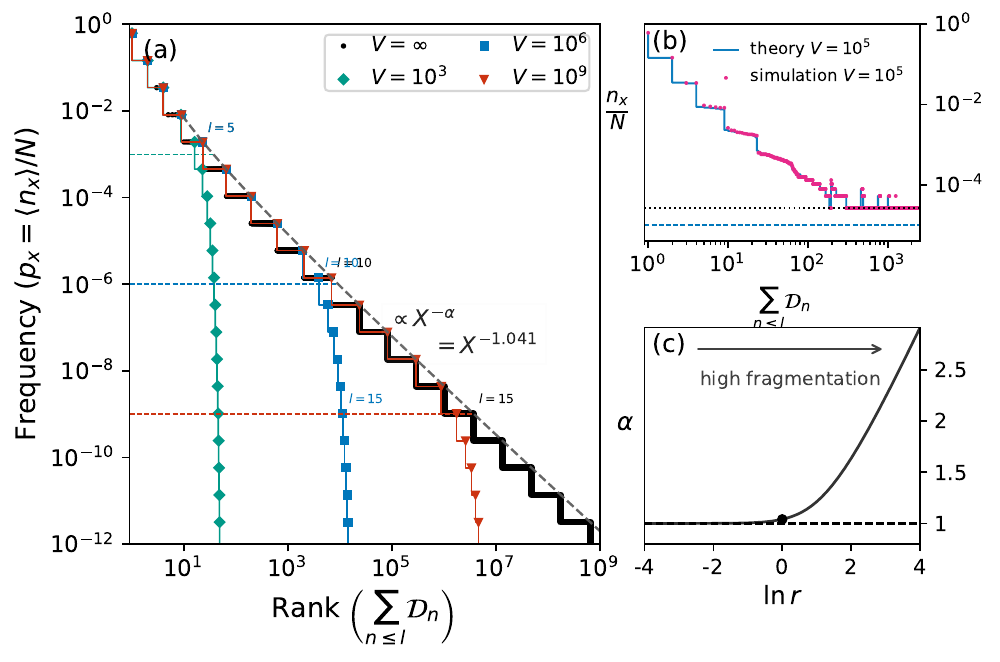}
		\caption{Rank-frequency representation of assemblies. \textbf{(a)} Exact equilibrium law $p_l$ at $r=1$, plotted against the cumulative diversity $\sum_{n\le l}\mathcal{D}_n$. The black staircase is the infinite-volume limit, where $\mathcal{D}_l=\mathcal{N}_l$, and the colored curves show the finite-size cumulative diversity for $V=10^3,10^6,10^9$. The dashed guide marks the scale $1/V$, and the gray line is the Zipf-like asymptotic $p_l\propto x^{-\alpha}$. \textbf{(b)} Comparison between finite-size Poisson theory and a Gillespie snapshot at $V=10^5$ and $r=1$, in terms of the normalized occupancy $n_x/N$, with dashed and dotted guides at $1/V$ and $1/(V\rho)$, respectively. \textbf{(c)} Dependence of the exponent $\alpha$ on the thermodynamic control parameter $\ln r$.}
		\label{fig:cumulative-diversity-rank}
	\end{figure}
	
	Consider first the infinite-volume system. At equilibrium, a single assembly of length $l$ has probability
	\begin{equation}
		p_l = \frac{c_l}{\rho} = \frac{r}{\rho}\, q^l ,
	\end{equation}
	and its rank, the cumulative number of assemblies of length at most $l$, is
	\begin{equation}
		X_l = \sum_{k\le l}\mathcal{N}_k .
	\end{equation}
	Since all assemblies of the same length share the same probability, each length contributes a plateau of width $\mathcal{N}_l$ in the rank-frequency curve, occupying the ranks between $X_{l-1}+1$ and $X_l$. This exact staircase structure is the black curve in Fig.~\ref{fig:cumulative-diversity-rank}(a); the individual plateaus are visible for small $l$ and merge into a smooth curve for larger $l$.
	
	In a finite-volume pool the accessible rank is no longer the cumulative count of possible assemblies but the cumulative count of realized ones,
	\begin{equation}
		X_l^{(V)} = \sum_{k\le l}\mathcal{D}_k .
	\end{equation}
	For $l\ll l_c$, $X_l^{(V)}\simeq X_l$. Beyond the crossover the accessible rank falls below $X_l$, because only a small fraction of the possible assemblies is realized. This finite-$V$ behavior is shown in Fig.~\ref{fig:cumulative-diversity-rank}(a) for $V=10^3,10^6,10^9$, with the black staircase being the infinite-volume limit $X_l^{(\infty)}=X_l$. The curves are progressively truncated once the average probability of a single assembly reaches $1/V$, while in an actual snapshot the minimal nonzero frequency is $1/N\approx1/(V\rho)$.
	
	To extract the large-rank asymptotics in the deterministic limit, the Catalan asymptotic gives
	\begin{equation}
		X_l \sim \frac{4}{3}\,\mathcal{N}_l \sim \frac{4^l}{3\sqrt{\pi}\,l^{3/2}} .
	\end{equation}
	The factor $4/3$ comes from last-term dominance of the cumulative sum: $\mathcal{N}_{l-k}/\mathcal{N}_l \to 4^{-k}$ for fixed $k$, hence $X_l/\mathcal{N}_l \to \sum_{k\ge0}4^{-k} = 4/3$. Using $q^l = 4^{-\alpha l}$ with $\alpha$ given by Eq.~\eqref{eq:alpha-def}, and inverting $X_l(l)$ at leading order (so that $l \sim \ln X_l/\ln 4$), one obtains
	\begin{equation}
		p_l = \frac{r}{\rho}\,4^{-\alpha l} \sim \frac{r}{\rho}\,(3\sqrt{\pi})^{-\alpha}\,X_l^{-\alpha}\,l^{-3\alpha/2} \sim A\,X_l^{-\alpha}\,(\ln X_l)^{-3\alpha/2} ,
	\end{equation}
	for a positive constant $A$. Hence, in the deterministic regime, the single-assembly probability is asymptotically Zipf-like in rank, with a logarithmic correction inherited from Catalan growth. In the coagulation-dominated limit $q\to 1/4$, $\alpha\to 1$ and the law approaches the canonical Zipf exponent.
	This mechanism is distinct from minimal dynamical models that generate Zipf-like laws by imposing a specific growth rule. In Simon's model, new types are introduced while existing types are copied with probability proportional to their current abundance, producing a Yule--Simon tail \cite{simon1955skew}. In proportional-growth models of city sizes, a Zipf distribution follows from Gibrat-like multiplicative growth together with stationarity conditions \cite{gabaix1999zipf}. In the present assembly network there is no preferential attachment, innovation rule, or global optimization. The elementary moves are local coagulation and fragmentation reactions satisfying detailed balance. The Zipf-like law instead appears at equilibrium from the exploration of the combinatorial assembly space: the exponential Gibbs weight of each individual assembly is balanced against the exponentially growing Catalan number of possible assemblies. The rank-frequency scaling is therefore the thermodynamic imprint of a combinatorial constraint on a simple reversible dynamics, rather than an ad hoc dynamical rule imposed to produce Zipf behavior.
	
	\section{Network Thermodynamics}
	\label{sec:thermodynamics}
	
	In Sec.~\ref{sec:equilibrium} the equilibrium concentrations $c_l=rq^l$ emerged as the solution of local detailed balance under mass conservation. We now show that this is a genuine thermodynamic equilibrium and extract the thermodynamics it carries. Two ingredients control it. First, the pool is in contact with a heat bath at temperature $T$: each internal bond carries an energy $\varepsilon_\mathrm{B}$, and local detailed balance makes the kinetic ratio $k_F/k_K$ exactly the Boltzmann weight for forming one bond. Second, the total number of atoms is conserved, and fixing it introduces an atomic chemical potential $\mu_\cS$ as the Lagrange multiplier conjugate to the atom count.
	
	These two ingredients act on different scales. The bond energy $\varepsilon_\mathrm{B}$ is \emph{local}, set by a single reaction channel. The chemical potential $\mu_\cS$ is \emph{global}: it is fixed only after summing over the entire Catalan assembly space at unit atom density, and so inherits the combinatorics. Once both are identified (Sec.~\ref{subsec:bond-chemical}), the factorized equilibrium measure becomes the grand-canonical ensemble of the closed pool, from which we obtain the bond density, energy, mean length, their fluctuations, and the entropy (Secs.~\ref{subsec:grand-canonical}--\ref{subsec:thermal-bath}). The entropy reappears as a complexity observable in Sec.~\ref{sec:complexity-measures}; here we only establish the thermodynamic potential from which it follows.
	
	\subsection{Bond Energy and Chemical Potential}\label{subsec:bond-chemical}
	
	At equilibrium the probability distribution factorizes over assembly types Eq.~\eqref{eq:equilib_dist}, so each type can be treated as thermodynamically independent of the rest even though the underlying pool is closed and atom conserving. The description is then grand-canonical and reservoir-like, with the fixed-mass pool recovered by conditioning the product-Poisson measure on the mass shell of Eq.~\eqref{eq:mass-conservation}, a standard construction for dilute reaction networks~\cite{SchmiedlSeifert2007}.
	
	We therefore write each equilibrium concentration in Gibbs form,
	\begin{equation}\label{eq:Gibbs}
		c_x = e^{-\beta \varphi(x)},
		\qquad
		\varphi(x)=\epsilon(x)-\mu(x),
	\end{equation}
	where $\varphi(x)$ is the \emph{one-copy free energy}: the thermodynamic cost of inserting a single copy of $x$, with internal energy $\epsilon(x)$ and chemical potential $\mu(x)$. It is distinct from the assembly grand potential $\Omega_x$ of Sec.~\ref{subsec:grand-canonical}, which appears only after summing over occupation numbers.
	
	No external reservoir is attached to each assembly type. Instead $\mu(x)$ is self-induced: through the reversible reactions, the rest of the pool acts as a structured reservoir of atoms and assemblies. Notice that only the combination $\varphi(x)=-T\ln c_x$ is fixed, by the equilibrium distribution and the network combinatorics, while the split into $\epsilon(x)$ and $\mu(x)$ is a convention, since only their difference enters Eq.~\eqref{eq:Gibbs}. We adopt the split with the clearest physical meaning: $\epsilon(x)$ records the heat exchanged with the bath when bonds are formed, and $\mu(x)$ records the conserved atomic material needed to build the assembly. 
	
	In logarithmic form, local detailed balance Eq.~\eqref{eq:local-db} gives one condition per reaction,
	\begin{equation}
		\varphi(x(y)) - \varphi(x) - \varphi(y) = -T\ln\!\left(\frac{k_K}{k_F}\right) ,
	\end{equation}
	fixing only the non-conserved increment. Because the network is loop-free, an assembly of length $l$ is built from $l$ monomers in exactly $l-1$ coagulations, so $\varphi$ is at most linear in length. And in particular
	\begin{equation}
		-\beta\varphi(x) = (l(x)-1)\ln \!\left(\frac{k_K}{k_F}\right) + l(x) \ln c_\cS ,
	\end{equation}
	in view of the explicit solution Eq.~\eqref{eq:equil-p_l}. The physical split is now set: each bond carries energy $\varepsilon_\mathrm{B}$, and the remaining length-proportional part is the chemical term,
	\begin{equation}
		\epsilon(x) = \varepsilon_\mathrm{B}\bigl(l(x)-1\bigr) ,
		\qquad
		\mu(x)=l(x)\,\mu_\cS ,
		\label{eq:en_mu}
	\end{equation}
	with 
	\begin{equation}\label{eq:energy}
		\varepsilon_\mathrm{B} =-T\ln\!\left(\frac{k_K}{k_F}\right) ,
		\qquad
		\mu_\cS = \beta^{-1}\ln c_\cS ,
	\end{equation} respectively the bond energy and the atomic chemical potential. Local detailed balance fixes the bond increment $\varepsilon_\mathrm{B}$, while the global mass constraint (i.e. the competition among all assemblies) fixes $\mu_\cS$ and therefore inherits the Catalan combinatorics. The kinetic parameters of Sec.~\ref{subsec:generating-function} can then be rewritten directly in thermodynamic form,
	\begin{equation}
		q = e^{-\beta(\varepsilon_\mathrm{B}-\mu_\cS)} ,
		\qquad
		r = e^{\beta\varepsilon_\mathrm{B}} .
	\end{equation}
	In view of this correspondence between kinetics and thermodynamics in the following we will use $\beta \varepsilon_\mathrm{B}$ instead of $\ln r$ as the external parameter.
	
	\subsection{Grand Canonical Potential}
	\label{subsec:grand-canonical}
	Each assembly type $x$ contributes an independent partition function, summing over occupation numbers with the classical indistinguishability weight\footnote{More generally, one may include a kinetic factor $K(x)$ in the single-assembly statistical weight, so that $p_x\propto K(x)e^{-\beta\varphi(x)}$. Throughout this paper we set $K(x)=1$, treating assemblies as chemical reactants in solution. For a classical ideal gas in $d$ dimensions, $K(x)=\lambda_{\mathrm{th}}(x)^{-d}$ with $\lambda_{\mathrm{th}}=h/\sqrt{2\pi m(x)/\beta}$ the thermal de Broglie length, adding a kinetic-energy contribution $\frac{d}{2}\ln(m(x)\beta^{-1}/2\pi)$ to the one-copy free energy.},
	\begin{equation}
		Z_x = \sum_{n_x\geq 0}\frac{\bigl(V\,e^{-\beta \varphi(x)}\bigr)^{n_x}}{n_x!} = \exp\bigl(V\,e^{-\beta \varphi(x)}\bigr) = e^{V c_x} ,
	\end{equation}
	where we used $\varphi(x) = \epsilon(x) - \mu(x) = -T\ln c_x$ from Eq.~\eqref{eq:Gibbs}. This is where the distinction between $\varphi(x)$ and $\Omega_x$ becomes concrete: $\varphi(x)$ is a one-copy free energy, while $\Omega_x$ is the grand-potential contribution obtained after summing over all possible copy numbers of that assembly with the associated  indistinguishability factorial,
	\begin{equation}
		-\beta\Omega_x = \ln Z_x = V c_x = \langle n_x \rangle ,
		\qquad
		\Omega_x=-T\langle n_x\rangle ,
	\end{equation}
	with $\langle\cdot\rangle$ denoting the equilibrium expectation.
	
	In this model the one-copy free energy of each assembly depends only on its length (as established in Sec.~\ref{subsec:equilibrium-solution}), since $\epsilon(x) = \varepsilon_\mathrm{B}(l(x)-1)$ and the equilibrium chemical potential is $\mu(x)=l(x)\mu_\cS$. The mean copy number reads $\langle n_l\rangle = V c_l = V r q^l$, and the total grand potential collapses to a length-weighted sum,
	\begin{equation}
		-\beta\Omega = \sum_x \langle n_x\rangle = \sum_{l\geq 1}\mathcal{N}_l \langle n_l\rangle = V r\, G_\mathcal{N}(q) ,
	\end{equation}
	using the generating function of Eq.~\eqref{eq:gen-func} and the substitutions $r = e^{\beta\varepsilon_\mathrm{B}}$, $q = e^{-\beta(\varepsilon_\mathrm{B} - \mu_\cS)}$. Since $\langle N\rangle = -\beta\Omega$, the total assembly density is $\rho = r\, G_\mathcal{N}(q)$, in agreement with Eq.~\eqref{eq:avg-length}.
	
	Within this thermodynamic description, the bond energy $\varepsilon_\mathrm{B}$ is fixed by the kinetics through Eq.~\eqref{eq:energy}, while the atomic chemical potential $\mu_\cS$ is determined globally by the unit-atom-density condition. Denoting the conserved total atom count by $M=\sum_x l(x)n_x$, this condition reads
	\begin{equation}
		\langle M\rangle = -\partial_{\mu_\cS}\Omega = V r q\, G_\mathcal{N}'(q) = V ,
	\end{equation}
	which reproduces Eq.~\eqref{eq:consV}. Solving gives
	\begin{equation}
		\mu_\cS = \varepsilon_\mathrm{B} - T\ln\!\bigl(2 + \sqrt{4 + e^{2\beta\varepsilon_\mathrm{B}}}\bigr) .
	\end{equation}
	
	\subsection{Thermodynamic Observables from the Grand Potential}
	
\begin{figure}[t]
		\centering
		\includegraphics[width=\columnwidth]{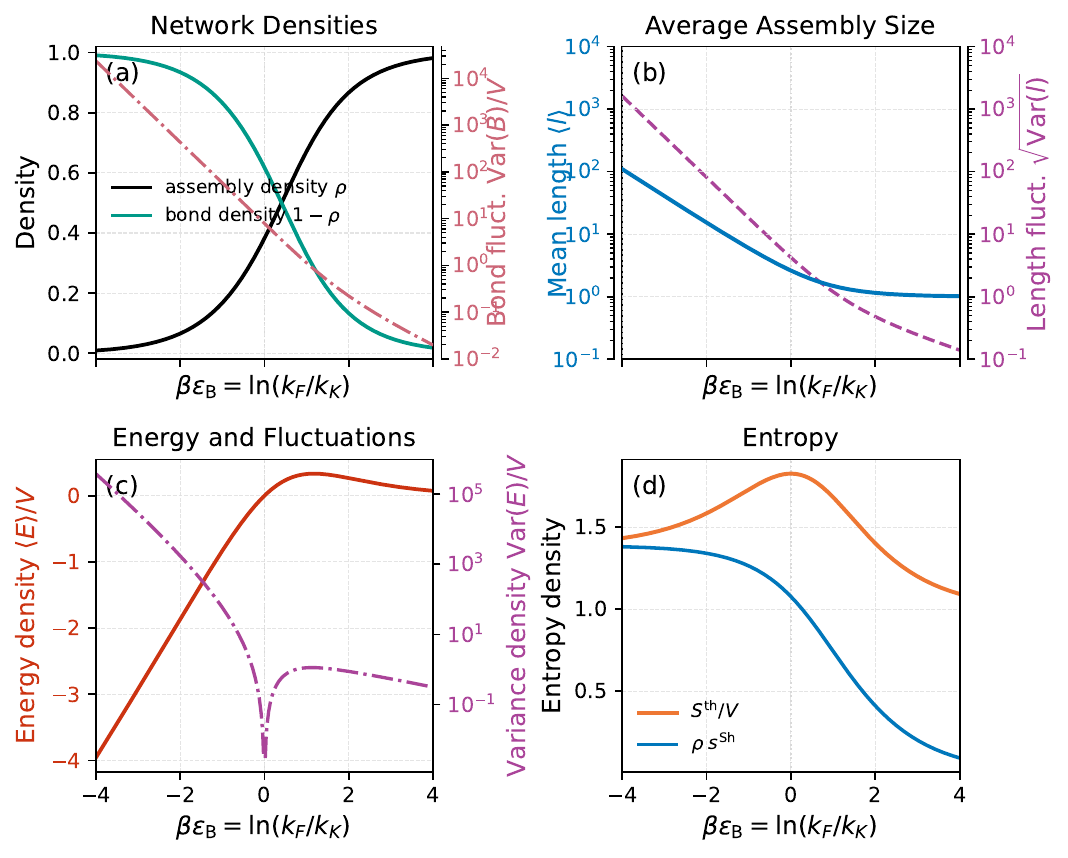}
		\caption{Thermodynamic observables derived from the constrained free energy and plotted versus the control parameter $\beta\varepsilon_\mathrm{B}=\ln(k_F/k_K)$. \textbf{(a)} Assembly density $\rho$, bond density $1-\rho$, and bond fluctuations $\mathrm{Var}(B)/V$. \textbf{(b)} Mean assembly length $\overline{l}=1/\rho$ and length fluctuations $\sqrt{\mathrm{Var}(l)}$, plotted on the same logarithmic range. \textbf{(c)} Energy density $\langle E\rangle/V$ and energy fluctuations $\mathrm{Var}(E)/V$. \textbf{(d)} Thermodynamic entropy density $S^{\mathrm{th}}/V$ and single-assembly Shannon entropy written as a density, $\rho s_{\rm exp}=Ns_{\rm exp}/V$. Because in this model each internal bond contributes one unit of energy $\varepsilon_\mathrm{B}$, the energetic observables are directly proportional to the corresponding bond observable, although this proportionality is special to the present network.}
		\label{fig:thermodynamic-observables}
	\end{figure}

	For this model a natural structural observable is the total number of internal bonds (parenthesis pairs),
	\begin{equation}
		\langle B \rangle
		= \left.\partial_{\varepsilon_\mathrm{B}}\Omega\right|_{\mu_\cS}
		=\sum_x \left\langle\bigl(l(x)-1\bigr)n_x \right\rangle
		= V(1-\rho)
		= V r\!\left[q G_\mathcal{N}'(q)-G_\mathcal{N}(q)\right]
	\end{equation}
	which counts the total number of binary coagulation events stored in the pool. Since the energy is proportional to the number of bonds, the mean thermal energy is
	\begin{equation}
		\langle E\rangle
		=\varepsilon_\mathrm{B}\langle B \rangle
		= \varepsilon_\mathrm{B} V(1-\rho) .
		\label{eq:appendix-E-mean}
	\end{equation}
	The bond fluctuations follow from the corresponding susceptibility,
	\begin{equation}
		\mathrm{Var}(B)
		= -T\left.\partial^2_{\varepsilon_\mathrm{B}}\Omega\right|_{\mu_\cS}
		= V r\!\left[
		q^2 G_\mathcal{N}''(q)-q G_\mathcal{N}'(q)+G_\mathcal{N}(q)
		\right].
	\end{equation}
	Since $E=\varepsilon_\mathrm{B} B$,
	\begin{equation}
		\mathrm{Var}(E)
		=\varepsilon_\mathrm{B}^2\mathrm{Var}(B).
		\label{eq:appendix-E-susc}
	\end{equation}
	The entropy follows from the use of the Legendre transform $\Omega = \langle\varphi\rangle - T S^{\mathrm{th}}$, which gives
	\begin{equation}
		T S^{\mathrm{th}} = \langle\varphi\rangle-\Omega = \langle E\rangle-\mu_\cS \langle M \rangle +T\langle N\rangle = V\left(\varepsilon_\mathrm{B}(1-\rho)-\mu_\cS+T\rho \right)
	\end{equation}
	which is the thermodynamic entropy of the pool, $S^{\mathrm{th}}_{\mathrm{pool}}$, discussed and compared with other entropy definitions in Sec.~\ref{subsec:entropy}. Equivalently we can rewrite it as a sum over species
	\begin{equation}
		S^{\mathrm{th}} = V\sum_x c_x\left[\beta\varphi(x)+1\right] = 
		V \sum_x c_x\left[\beta(\epsilon(x)-\mu(x))+1\right] = V \sum_x c_x\left[1-\ln c_x\right] .
		\label{eq:entropy-sum-species}
	\end{equation}
	which will be the description adopted when dealing with the transient dynamics in a paper in preparation~\cite{FolenaKruszewski2026FreeExpansion}.
	Finally, the length of a randomly chosen realized assembly is distributed with probabilities $p_x=c_x/\rho$. Its mean is therefore
	\begin{equation}
		\overline l = \frac{\langle M \rangle}{\langle N\rangle} = \frac{V r q G_\mathcal{N}'(q)}{\langle N\rangle} = \frac{1}{\rho} .
	\end{equation}
	The corresponding variance is
	\begin{equation}
		\mathrm{Var}(l) = \sum_x p_x l(x)^2-\overline l^{\,2}
		= \frac{r}{\rho}\left[
		q G_\mathcal{N}'(q)+q^2G_\mathcal{N}''(q)
		\right]
		-\frac{1}{\rho^2}.
	\end{equation}
	Substituting $q(\varepsilon_\mathrm{B}) = 1/(2+\sqrt{4+e^{2\beta\varepsilon_\mathrm{B}}})$ makes every observable an explicit function of the control parameter $\beta\varepsilon_\mathrm{B} = \ln(k_F/k_K)$; Fig.~\ref{fig:thermodynamic-observables} summarizes the results.
	
	\subsection{Thermal Bath}
	\label{subsec:thermal-bath}
	
	All equilibrium formulae depend on the single dimensionless control parameter $u=\beta\varepsilon_\mathrm{B}$, or equivalently on $r=e^u$. One can therefore either set $T=1$ and vary the signed bond energy $\varepsilon_\mathrm{B}=u$, as in Fig.~\ref{fig:thermodynamic-observables}, or fix a bond-energy scale and vary the temperature. If the heat bath is restricted to positive temperatures, the latter convention splits the model into two physical branches.
	
	For a positive bond energy, $\varepsilon_\mathrm{B}=+1$, bond formation is energetically costly and the system is fragmenting. The high-temperature limit gives $u=1/T\to0$, hence $r\to1$, the \emph{equal-rate point} ($k_F=k_K$): the infinite-temperature point, or equivalently the entropy-driven point, where no energy is attributed to bonds ($\beta\varepsilon_\mathrm{B}=0$). The low-temperature limit gives $u\to+\infty$, hence $r\to\infty$, $\rho\to1$, and the pool becomes monomer dominated. For a negative bond energy, $\varepsilon_\mathrm{B}=-1$, bond formation is favorable and the system is coagulating. Again $T\to\infty$ gives the same equal-rate point, but now $T\to0^+$ gives $r=e^{-1/T}\to0$ and drives the pool toward the full-coagulation limit. This mirroring around the infinite-temperature equal-rate point is the reason the entropy is maximal at
	$\ln r = 0$.
	
	Near the equal-rate point $u=0$, the generating function is evaluated away from its Catalan singularity and all thermodynamic densities vary smoothly. The singular finite-size issue appears instead on the low-temperature coagulating branch, where $r\ll1$, $q\to1/4$, and the length distribution approaches the Catalan radius of convergence. In that regime one has $\rho\simeq r/2$ and $\overline{l}\simeq2r^{-1}$. Equivalently, for $\varepsilon_\mathrm{B}=-1$ at low temperature,
	\begin{equation}
		\rho\simeq \frac{1}{2}e^{-1/T},
		\qquad
		\overline{l}\simeq 2e^{1/T}.
	\end{equation}
	The dominant fluctuations grow as\footnote{On the coagulating branch $r=e^{-1/T}\to0$, so $q=1/(2+\sqrt{4+r^2})\to1/4$ with $s\equiv\sqrt{1-4q}\simeq r/4$. Writing $G_\mathcal{N}(q)=(1-s)/2$ gives $G_\mathcal{N}'(q)=s^{-1}$ and $G_\mathcal{N}''(q)=2s^{-3}$, so the leading term of each susceptibility is $q^2G_\mathcal{N}''\simeq\tfrac18 s^{-3}=8/r^3$ (while $qG_\mathcal{N}'\simeq1/r$ and $G_\mathcal{N}\simeq\tfrac12$). Hence $\mathrm{Var}(B)/V=r\,[q^2G_\mathcal{N}''-qG_\mathcal{N}'+G_\mathcal{N}]\simeq8/r^2=8e^{2/T}$, and, using $\rho\simeq r/2$ so that $r/\rho\simeq2$, $\mathrm{Var}(l)=\tfrac{r}{\rho}[qG_\mathcal{N}'+q^2G_\mathcal{N}'']-\rho^{-2}\simeq16/r^3=16e^{3/T}$; the energy follows from $\mathrm{Var}(E)=\varepsilon_\mathrm{B}^2\,\mathrm{Var}(B)$.}
	\begin{equation}
		\frac{\mathrm{Var}(B)}{V}\simeq 8 e^{2/T},
		\qquad
		\frac{\mathrm{Var}(E)}{V}\simeq 8\varepsilon_\mathrm{B}^2 e^{2/T},
		\qquad
		\mathrm{Var}(l)\simeq 16 e^{3/T}.
	\end{equation}
	Thus the energy and length fluctuations grow exponentially as the network approaches the full-coagulation limit. The standard deviation $\sqrt{\mathrm{Var}(l)}\propto e^{3/(2T)}$ grows faster than the typical length $\overline{l}\propto e^{1/T}$, so relative size fluctuations diverge and the pool becomes extremely heterogeneous. At finite volume this growth competes with the finite-size scales of Fig.~\ref{fig:length-scales}(b): once the mean assembly number drops to $V\rho=O(1)$, the conserved mass can no longer be shared among many molecules and a single length-$O(V)$ assembly captures a finite fraction of all atoms. The pool then undergoes a crossover condensation in temperature, setting in on the $\varepsilon_\mathrm{B}=-1$ branch below $T_c\sim 1/\ln V$. This is the  finite-volume counterpart of the condensation found in balls-in-boxes process~\cite{Bialas2023}. It remains a genuine finite-size effect: $T_c$ drifts logarithmically to zero, so no condensate survives at any fixed temperature as $V\to\infty$.
	
	\section{Measures of Complexity}
	\label{sec:complexity-measures}
	
	We now turn from the thermodynamic description to observables that quantify how much structural complexity is visible in a finite pool. Thermodynamic entropy alone does not capture this: it is insensitive to whether the populated states contain long or varied assemblies. For example, a pool containing two distinguishable length-$10$ assemblies and a pool containing one length-$1$ plus one length-$19$ assembly have the same mass ($20$), the same number of bonds ($18$, hence the same energy), and the same number of assemblies (two). They are thermodynamically indistinguishable at this level, even though different structural observables would rank them differently.
	
	We therefore use ``complexity'' as a family of observables, not as a single thermodynamic quantity. The \emph{diversity} $\mathcal{D}$ counts how many distinct types are visible; the \emph{Shannon entropies} measure how heterogeneous the equilibrium distribution or the occupation-number ensemble is; and \emph{assembly-weighted} observables $\mathcal{A}$ give extra weight to occupied long hierarchical structures. The last family is adapted from assembly measures introduced to quantify how far a chemical soup has been driven by selection~\cite{MarshallEtAl2021,sharma2023assembly}. In the present model its length bias is also thermodynamically interpretable: it is equivalent to a uniform shift of the bond energy (Sec.~\ref{subsec:assembly-complexity}).
	
	Our reversible network is deliberately not a selective system. It has no catalysis, no functional target, and no external growth rule; all structure arises from detailed balance on a combinatorially expanding reaction graph. The observables that we consider define a \emph{neutral baseline}: any future selective or irreversible chemistry should be compared against this baseline before structural growth is attributed to selection.
	
\begin{figure*}[t]
		\centering
		\includegraphics[width=0.98\textwidth]{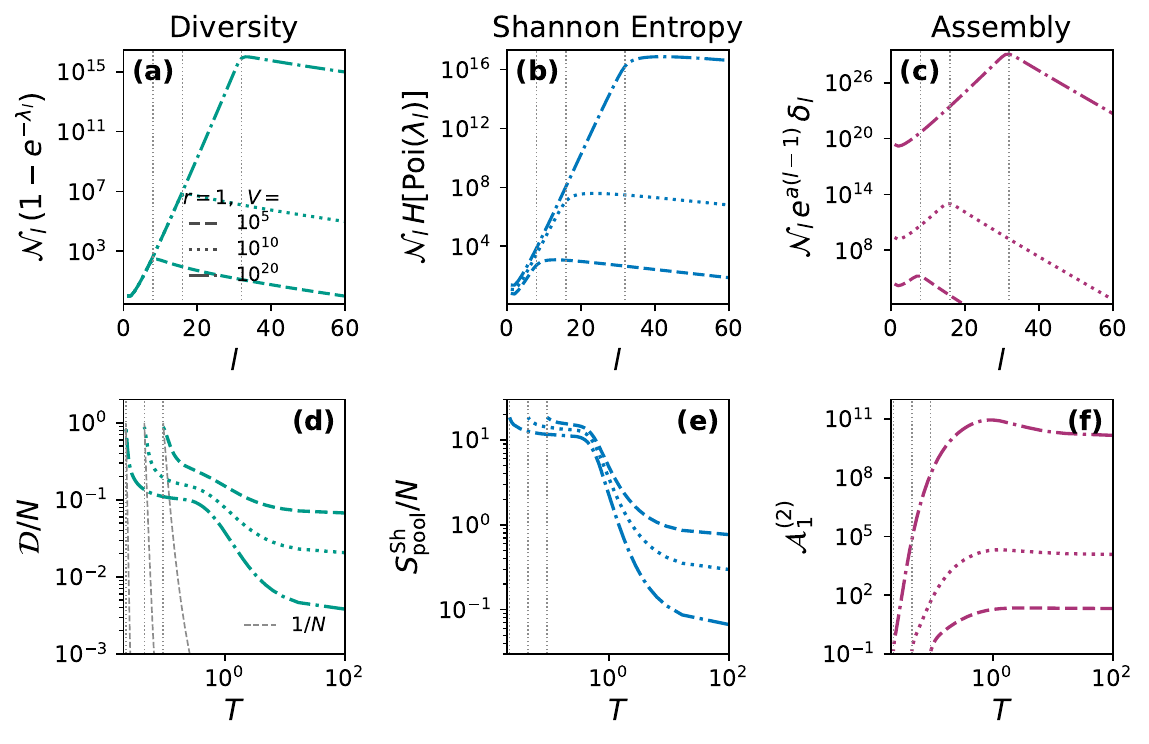}
		\caption{Finite-volume complexity observables, for $V=10^5,10^{10},10^{20}$ (dashed, dotted, dash-dotted). \textbf{(a--c) length dependence at $r=1$.} Each observable is a shell sum $\mathcal{O}=\sum_l\mathcal{N}_l\,w_l\,\phi(\lambda_l)$ [Eq.~\eqref{eq:master-sum}]; its summand vs.\ length $l$ forms a hump on the crossover shell $l_c=\ln(Vr)/\ln(1/q)$ (dotted verticals), where $\lambda_l=Vrq^l=O(1)$: (a) diversity, (b) occupation Shannon entropy $S^\mathrm{Sh}_{\mathrm{pool}}$, (c) assembly ($a=1$). \textbf{(d--f) Temperature dependence}, coagulating branch $\varepsilon_\mathrm{B}=-1$ ($r=e^{-1/T}$); dotted verticals mark the temperature $T_c$ of the condensation crossover $N=V\rho=1$. (d) diversity per assembly $\mathcal{D}/N$ and floor $1/N$. (e) Shannon entropy of the pool per assembly $S^{\mathrm{Sh}}_{\mathrm{pool}}/N$, the integrated counterpart of the spectrum in (b) (drawn only while $N=V\rho\geq1$); it grows as the pool dilutes on cooling. (f) assembly $\mathcal{A}^{(2)}_1$. All three lower panels are per-assembly ratios ($1/N=1/(V\rho)$), a normalization that lowers the volume exponent by one relative to the total observable (Table~\ref{tab:complexity-scaling}); diversity and Shannon entropy are subextensive, while the assembly is superexpensive.}
		\label{fig:complexity-temperature}
	\end{figure*}

	All these observables can be written as a sum over length shells
	\begin{equation}
		\mathcal{O}=\sum_{l\ge1}\mathcal{N}_l\,w_l\,\phi(\lambda_l),
		\qquad \lambda_l=V c_l=Vr\,q^l ,
		\label{eq:master-sum}
	\end{equation}
	with an observable-dependent length weight $w_l$ and occupation factor $\phi$. The Catalan multiplicity $\mathcal{N}_l$ grows asymptotically as $\sim4^l/l^{3/2}$, while the mean occupancy $\lambda_l=\langle n_l\rangle$ decays geometrically as $rq^{l}$. The crossover shell satisfies
	\begin{equation}
		\lambda_{l_c}=O(1),
		\qquad
		l_c=\frac{\ln(Vr)}{\ln(1/q)}
	\end{equation}
	at fixed $r$. For every observable of the form~\eqref{eq:master-sum}, most of the contribution comes from lengths near $l_c$. The measures differ mainly through $w_l$, and therefore through a single size-scaling exponent $\delta$ (Sec.~\ref{subsec:master-scaling}, Table~\ref{tab:complexity-scaling}). The top row of Fig.~\ref{fig:complexity-temperature}(a--c) shows the same hump at $l_c$ for the diversity, the pool Shannon entropy, and assembly-weighted observables.
	
	\subsection{Diversity}
	\label{subsec:diversity-complexity}
	
	Diversity counts how many distinct assemblies are present in the pool. It is not a structural complexity measure by itself, because it is blind to the length and shape of the assemblies. It is nevertheless the most direct finite-volume measure of how much of the combinatorial species space has been realized.
	
	Summing the length-resolved diversity of Eq.~\eqref{eq:diversity-single} gives the total number of distinct realized assemblies,
	\begin{equation}
		\mathcal{D}(V)
		= \sum_{l\ge1}\mathcal{D}_l
		= \sum_{l\ge1}\mathcal{N}_l\bigl(1-e^{-\langle n_l\rangle}\bigr).
	\end{equation}
	Its large-volume scaling follows from a saddle-point estimate with $l$ taken continuous. Writing the summand as $\mathcal{D}_l=e^{f(l)}$, with $\mathcal{N}_l\sim 4^{l-1}/[\sqrt\pi\,(l-1)^{3/2}]$ [Eq.~\eqref{eq:catalan-count}] and $\lambda_l\equiv\langle n_l\rangle=Vr\,4^{-\alpha l}$ [$q=4^{-\alpha}$, Eq.~\eqref{eq:alpha-def}],
	\begin{equation}
		f(l) = l\ln4 - \tfrac32\ln l + \ln\!\bigl(1-e^{-\lambda_l}\bigr) + \mathrm{const} ,
	\end{equation}
	the sum is dominated by its maximum,
	\begin{equation}
		\mathcal{D}(V)\simeq\int\!\mathrm{d}l\;e^{f(l)} \approx e^{f(l^*)},\qquad f'(l^*)=0 .
		\label{eq:diversity-laplace}
	\end{equation}
	The growing Catalan term $l\ln4$, balanced against the occupancy cutoff $\ln(1-e^{-\lambda_l})$, places the maximum on the crossover shell where $\lambda_{l^*}=O(1)$. To make this explicit, we rescale the length by the crossover scale $l_c=\ln(Vr)/(\alpha\ln4)$, setting $\hat l=l/l_c$; the two length-dependent factors then become pure powers of $Vr$,
	\begin{equation}
		4^{l}=(Vr)^{\hat l/\alpha},
		\qquad
		\lambda_l=Vr\,4^{-\alpha l}=(Vr)^{1-\hat l},
	\end{equation}
	so the mean occupancy $\lambda_l$ crosses unity precisely at $\hat l=1$. This crossing is what splits the occupancy factor into two regimes,
	\begin{equation}
		\ln\!\bigl(1-e^{-\lambda_l}\bigr)\simeq
		\begin{cases}
			0, & \hat l<1\quad(\lambda_l\gg1,\ \text{saturated}),\\
			\ln\lambda_l=(1-\hat l)\ln(Vr), & \hat l>1\quad(\lambda_l\ll1,\ \text{rare}),
		\end{cases}
	\end{equation}
	and hence into the two branches of the rate function. Dropping the subexponential $-\tfrac32\ln l$ term, the exponent takes the scaling form $f\simeq\ln(Vr)\,\psi(\hat l)$, in which the Catalan growth $l\ln4=(\hat l/\alpha)\ln(Vr)$ adds to this occupancy contribution:
	\begin{equation}
		\psi(\hat l)=\frac{\hat l}{\alpha}+
		\begin{cases}
			0 \\
			1-\hat l
		\end{cases}
		\;=\;
		\begin{cases}
			\hat l/\alpha, & \hat l<1,\\
			1-\hat l\,(\alpha-1)/\alpha, & \hat l>1.
		\end{cases}
		\label{eq:diversity-tent}
	\end{equation}
	Below the crossover only the Catalan growth survives and $\psi$ rises with slope $1/\alpha$; above it the decaying occupancy takes over and $\psi$ falls with slope $-(\alpha-1)/\alpha$. The two branches therefore meet at the crossover shell $\hat l=1$, where $\psi$ peaks at $\psi(1)=1/\alpha$ [dotted vertical lines in Fig.~\ref{fig:complexity-temperature}(a)]. As $V\to\infty$ the prefactor $\ln(Vr)$ diverges, so $e^{f}$ concentrates on $\hat l=1$ with vanishing width $\Delta\hat l\sim1/\ln(Vr)$ and the saddle-point estimate becomes exact at leading exponential order. Restoring the $-\tfrac32\ln l$ prefactor at $l^*=l_c$ gives $e^{f(l^*)}\propto 4^{l_c}/l_c^{3/2}$ with $4^{l_c}=(Vr)^{1/\alpha}$ and $l_c\propto\ln(Vr)$,
	\begin{equation}
		\mathcal{D}(V)
		\propto
		\frac{(Vr)^{1/\alpha}}{[\ln(Vr)]^{3/2}} .
		\label{eq:diversity-scaling}
	\end{equation}
	The power $1/\alpha=\psi(1)$ is the Catalan growth cut off at the occupancy crossover, and the $[\ln(Vr)]^{-3/2}$ is the sub-exponential Catalan prefactor at that same shell. Since $\alpha>1$, the total diversity is subextensive in $V$. Up to the logarithmic correction this is a Heaps law: the number of observed types grows as a power of the number of tokens, with Heaps exponent $1/\alpha$ fixed by the inverse Zipf exponent of the rank-frequency law~\cite{lu2010zipf,MazzarisiEtAl2021}. The rare sector carries the same power, since $N\,P(L>l_c)\propto\mathcal{D}(V)$: the species beyond $l_c$ are as numerous as the entire occupied diversity. Finally, the ratio $\mathcal{D}(V)/N$ is the analogue of the type-token ratio used to measure linguistic diversity, which in large text corpora likewise decays as a power of the corpus size~\cite{rosillo2025}.
	
	Figure~\ref{fig:complexity-temperature}(d) shows the diversity density $\mathcal{D}/N$ as a function of temperature in the coagulating branch $\varepsilon_\mathrm{B}<0$ for three different volumes $V=10^5,10^{10},10^{20}$: it rises toward unity as the pool cools and meets its single-type floor $1/N$ at finite-size condensation $T_c$, where only a small number of large assemblies is present in the system (see Sec.~\ref{subsec:thermal-bath}). Notice that there are two sectors: for high temperature the three considered volumes are already in the asymptotic sub-extensive regime, therefore when rescaled by $N$ they collapse to zero; instead in the low temperature sector they are still in the pre-asymptotic regime and still scale well with $N$.
	
	In terms of the master shell sum Eq.~\eqref{eq:master-sum} diversity has no extra length bias: $w_l=1$, $\phi(\lambda_l)=1-e^{-\lambda_l}$ and its exponent is $\delta=1/\alpha$, or $\delta=1/\alpha-1$ after division by $N$, as reported by in Table~\ref{tab:complexity-scaling}.
	
	\subsection{Shannon Entropies}\label{subsec:entropy}
	
	Three entropies enter this discussion. The \emph{thermodynamic} entropy $S^{\mathrm{th}}$ is extensive and governs the thermodynamics laws. The \emph{single-assembly} entropy $s^{\mathrm{Sh}}\equiv-\sum_x p_x\ln p_x$ is intensive: it is the Shannon entropy of the normalized type distribution $p_x=c_x/\rho$. The \emph{pool} entropy $S^{\mathrm{Sh}}_{\mathrm{pool}}\equiv-\sum_{\mathbf n} \mathbb{P}_{\mathbf n}\ln\mathbb{P}_{\mathbf n}$ is the Shannon entropy of the full occupation-number ensemble. Despite the similar notation, it is neither $s^{\mathrm{Sh}}$ nor $S^{\mathrm{th}}$; in fact it is subextensive in $V$ (Appendix~\ref{app:shannon_pool}). In short, $S^{\mathrm{th}}$ is thermodynamic and extensive, $s^{\mathrm{Sh}}$ is type-distribution and intensive, and $S^{\mathrm{Sh}}_{\mathrm{pool}}$ is occupation-number and subextensive.
	
	The complexity-relevant entropies are the two Shannon entropies, not the thermodynamic entropy. The thermodynamic entropy is maximal at the equal-rate point $r=1$ [Fig.~\ref{fig:thermodynamic-observables}(d)], whereas $s^{\mathrm{Sh}}$ grows along the coagulating branch because the normalized distribution spreads over longer assemblies. Thus, for $\varepsilon_\mathrm{B}=-1$ and $T\to0$, $S^{\mathrm{th}}$ decreases while $s^{\mathrm{Sh}}$ increases. The pool entropy $S^{\mathrm{Sh}}_{\mathrm{pool}}$ behaves differently again: as a finite-volume occupation entropy, it is controlled by the same rare crossover shell that controls diversity.
	
	\subsubsection{Single Assembly Entropy}
	Both $s^{\mathrm{Sh}}$ and the thermodynamic entropy per assembly [Eq.~\eqref{eq:entropy-sum-species}] are set by the mean one-copy free energy $\overline{\varphi}=\sum_x p_x\varphi(x)$,
	\begin{equation}
		s^{\mathrm{Sh}}\equiv-\sum_x p_x\ln p_x=\ln\rho+\beta\overline{\varphi},
		\qquad
		\frac{S^{\mathrm{th}}}{N}=1+\beta\overline{\varphi}.
	\end{equation}
	In our model $\beta\overline{\varphi}=-\ln r-(\ln q)/\rho$, evaluated from $c_l=rq^l$ and $\overline{l}=1/\rho$ [Eq.~\eqref{eq:avg-length}]. Therefore
	\begin{equation}
		\rho s^{\mathrm{Sh}}=\frac{S^{\mathrm{th}}}{V}-\rho(1-\ln\rho).
		\label{eq:entropy-joint-per-assembly}
	\end{equation}
	Since $\rho\leq1$, the correction $\rho(1-\ln\rho)$ is non-negative and $S^{\mathrm{th}}/V\geq\rho s^{\mathrm{Sh}}$, with equality approached in the full-coagulation limit $\rho\to0$. Figure~\ref{fig:thermodynamic-observables}(d) plots these two densities against $\ln r=\beta\varepsilon_\mathrm{B}$: on the coagulating branch, cooling raises $\rho s^{\mathrm{Sh}}$ while lowering $S^{\mathrm{th}}/V$, and both meet at $-\lim_{\beta\to\infty}\ln q=\ln4$. On the fragmenting branch both densities decrease as the pool becomes monomer dominated. Notice however that if we normalize per assembly instead of per volume, $S^{\mathrm{th}}/N$ and $s^{\mathrm{Sh}}$ rise together on the coagulating branch and fall together on the fragmenting branch. 
	
	\subsubsection{Pool Entropy}
	The Shannon pool entropy is a shell-sum observable. Grouping the factorized grand-canonical occupation distribution by length gives
	\begin{equation}
		S^{\mathrm{Sh}}_{\mathrm{pool}}
		\equiv -\sum_{\mathbf n} \mathbb{P}_{\mathbf n}\ln\mathbb{P}_{\mathbf n}
		= \sum_x H\bigl(\mathrm{Poi}(\lambda_x)\bigr)
		= \sum_l \mathcal{N}_l\,H\bigl(\mathrm{Poi}(\lambda_l)\bigr),
		\label{eq:entropy-sh-pool-shells}
	\end{equation}
	where $\mathrm{Poi}(\lambda;n)\equiv e^{-\lambda}\lambda^{n}/n!$ and
	\begin{equation}
		H\bigl(\mathrm{Poi}(\lambda)\bigr)
		\equiv -\sum_{n\ge0} \mathrm{Poi}(\lambda;n)\ln \mathrm{Poi}(\lambda;n)
	\end{equation}
	is the Shannon entropy of one Poisson-distributed occupation number. This is Eq.~\eqref{eq:master-sum} with $w_l=1$ and $\phi(\lambda_l)=H(\mathrm{Poi}(\lambda_l))$: the same unit weight as the diversity, only with a different occupation factor. The Poisson entropy
	\begin{equation}
		H\bigl(\mathrm{Poi}(\lambda)\bigr)\simeq
		\begin{cases}
			\tfrac12\ln(2\pi e\lambda), & \lambda\gg1 \quad(\text{high-copy bulk}),\\[2pt]
			O(1), & \lambda\sim1 \quad(\text{crossover shell}),\\[2pt]
			-\lambda\ln\lambda, & \lambda\ll1 \quad(\text{rare tail}),
		\end{cases}
	\end{equation}
	is $O(1)$ on the crossover shell, only logarithmically large in the high-copy bulk below it, and vanishing in the rare tail above it. Its profile is therefore the same as that of the diversity factor $1-e^{-\lambda_l}$ up to a slowly varying logarithm, so, as for every observable of the form~\eqref{eq:master-sum}, the sum is dominated by the crossover shell $l_c$, where $\phi=O(1)$. Each of the $\mathcal{N}_{l_c}$ species visible there carries an $O(1)$ amount of entropy, so $S^{\mathrm{Sh}}_{\mathrm{pool}}$ counts essentially the same species as the diversity and inherits its exponent,
	\begin{equation}
		S^{\mathrm{Sh}}_{\mathrm{pool}}
		\propto \frac{(Vr)^{1/\alpha}}{[\ln(Vr)]^{3/2}} .
	\end{equation}
	
	\subsubsection{Empirical Entropy}
	There is one practical measurement caveat. Recovering $s^{\mathrm{Sh}}$ from a single finite pool snapshot $\mathbf n$ is difficult because the rare species near $l_c$ are precisely the ones most likely to be missed. The plug-in estimator
	\begin{equation}
		\widehat{s}^{\mathrm{Sh}}(\mathbf{n})=-\sum_{x:\,n_x>0}\frac{n_x}{N}\ln\!\left(\frac{n_x}{N}\right)
		\label{eq:entropy-snapshot-estimator}
	\end{equation}
	is therefore biased downward. The missing contribution is controlled by the same pool entropy,
	\begin{equation}
		s^\mathrm{Sh}-\mathbb{E}[\widehat{s}^{\mathrm{Sh}}(\mathbf{n})]
		\sim \frac{S^\mathrm{Sh}_\mathrm{pool}}{\rho V}
		\propto \frac{V^{1/\alpha-1}}{[\ln(V)]^{3/2}} .
	\end{equation}
	At $r=1$ the algebraic part is only about $V^{-0.04}$, and it becomes slower with stronger coagulation. Appendix~\ref{app:entropy-snapshot} gives the detailed estimator calculation.
	
	\subsection{Assembly}
	\label{subsec:assembly-complexity}
	
	Assembly-weighted observables add a structural bias to diversity by favoring longer assemblies. We adapt the assembly measures of Sharma \textit{et al.}~\cite{sharma2023assembly}, assigning each assembly $x$ a weight $e^{a(l(x)-1)}$, where $l(x)-1$ is the number of binary coagulation events recorded in $x$. In this model that count is also proportional to the bond energy, so the bias $a$ is thermodynamically equivalent to subtracting $\beta^{-1}a$ from each bond energy.
	
	Writing $\lambda_l=\langle n_l\rangle=Vrq^l$ and using the Poisson occupation mass $\mathrm{Poi}(\lambda_l;k)=e^{-\lambda_l}\lambda_l^k/k!$ of one length-$l$ assembly, we use two related observables,
	\begin{equation}
		\begin{aligned}
			\mathcal{A}^{(1)}_{a,l}
			={}& \frac{e^{a(l-1)}}{V\rho}\,
			\mathcal{N}_l
			\sum_{k\ge1}\mathrm{Poi}(\lambda_l;k)
			= \frac{e^{a(l-1)}}{V\rho}\,
			\mathcal{N}_l\bigl(1-e^{-\lambda_l}\bigr), \\
			\mathcal{A}^{(2)}_{a,l}
			={}& \frac{e^{a(l-1)}}{V\rho}\,
			\mathcal{N}_l
			\sum_{k\ge2}(k-1)\,
			\mathrm{Poi}(\lambda_l;k)
			= \frac{e^{a(l-1)}}{V\rho}\,
			\mathcal{N}_l\bigl(\lambda_l-1+e^{-\lambda_l}\bigr) .
		\end{aligned}
	\end{equation}
	Their totals are
	\begin{equation}
		\mathcal{A}^{(i)}_a(V)=\sum_{l\ge1}\mathcal{A}^{(i)}_{a,l},
		\qquad i=1,2 .
	\end{equation}
	Here $\mathcal{A}^{(1)}_{a,l}$ counts each realized assembly once, whereas $\mathcal{A}^{(2)}_{a,l}$ counts only copies beyond the first. Thus $\mathcal{A}^{(2)}$ is the Sharma-style excess-copy assembly observable, while $\mathcal{A}^{(1)}$ is an auxiliary occupied-species variant introduced for comparison. For $a=0$, the first reduces to the diversity density, $\mathcal{A}^{(1)}_0(V)=\mathcal{D}(V)/(V\rho)$, and the choice $a=1$ in the excess-copy variant $\mathcal{A}^{(2)}_a(V)$ reproduces the original linear assembly weight $e^{l-1}$ of \cite{sharma2023assembly}.
	
	These observables have again the same shell decomposition, but this time with an explicit weight $w_l= e^{a(l-1)}$, which induces a modified scaling (Appendix~\ref{app:assembly-weighted})
	\begin{equation}
		\mathcal{A}^{(1)}_a(V),\ \mathcal{A}^{(2)}_a(V)
		\propto
		\frac{(Vr)^{\delta_A(a)}}{[\ln(Vr)]^{3/2}},
	\end{equation}
	where the power law reads
	\begin{equation}
		\delta_A(a)
		=\frac{1}{\alpha}\Big(1+\frac{a}{\ln4}\Big)-1 = \frac{a-a_c^{(1)}}{-\ln q} .
	\end{equation}
	The length bias contributes $a/\ln4$ in units of the Catalan growth rate, and the $-1$ appears because these observables are normalized per assembly. The two definitions share this leading exponent wherever their series converge, but they have different convergence windows. The reason is the following: $\mathcal{A}^{(2)}_a$ counts only copies beyond the first, so its generating-function series starts one order later and remains finite for a larger range of length biases. This is what lets $\mathcal{A}^{(2)}$ reach positive $\delta_A$, so the normalized observable grows with $V$ and the corresponding unnormalized total is superextensive. The limiting values of $a$ follow from requiring the terms $G_{\mathcal N}(e^a q^m)$ of the series to remain within the Catalan radius of convergence $1/4$:
	\begin{equation}
		a_c^{(1)}=-\ln(4q),
		\qquad
		a_c^{(2)}=-\ln(4q^2)=a_c^{(1)}-\ln q.
	\end{equation}
	The difference arises because $\mathcal{A}^{(1)}_a$ starts at $m=1$, whereas for $\mathcal{A}^{(2)}_a$ the subtraction of the first copy removes that term and the series starts at $m=2$. The maximal power-law reachable by the first definition is $\delta_A(a_c^{(1)})=0$, while $\delta_A(a_c^{(2)})=1$. Hence $\mathcal{A}^{(2)}$ is the natural observable for exploring positive $\delta_A$, corresponding to superextensive scaling in $V$.
	
	A rescaled saddle-point analysis in the length $\hat l=l/l_c$ (Appendix~\ref{app:assembly-weighted}) shows that, as long as the exponent stays in its feasible window ($0<\delta_A<1$ for $\mathcal{A}^{(2)}$, and $-1<\delta_A<0$ for $\mathcal{A}^{(1)}$), the dominant shell remains the diversity crossover $l_c$, displaced only by an $O(1)$ shift toward longer assemblies; outside that window the peak detaches from $l_c$. The crossover \emph{scale} $l_c\sim\ln(Vr)$ is therefore unchanged by the assembly bias throughout the window.
	
\begin{figure*}[t]
		\centering
		\includegraphics[width=0.98\textwidth]{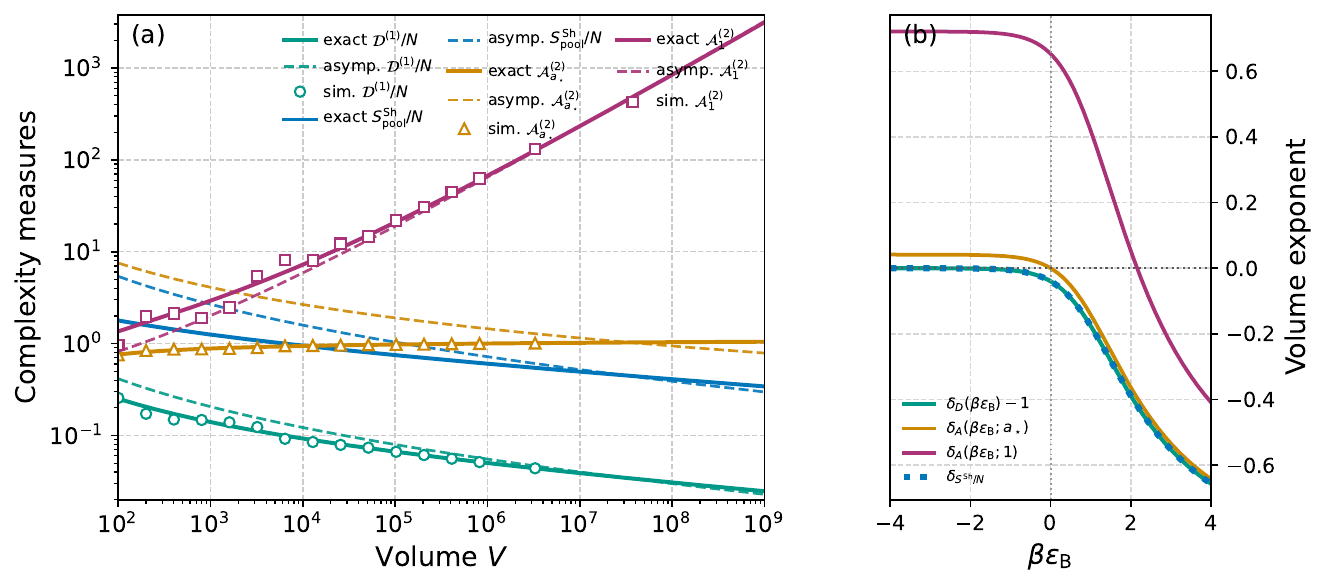}
		\caption{Representative normalized complexity observables at $r=1$. \textbf{(a)} Occupied-species density $\mathcal{D}(V)/N=\mathcal{D}(V)/(V\rho)$, pool Shannon entropy per assembly $S^{\mathrm{Sh}}_{\mathrm{pool}}/N$, marginal excess-copy observable $\mathcal{A}^{(2)}_{a_\star}(V)$ with $a_\star=a_c^{(1)}|_{r=1}$, and Sharma-style assembly observable $\mathcal{A}^{(2)}_1(V)$. Solid, dashed, and open-symbol curves denote the exact theory, the corresponding asymptotic forms, and Gillespie simulations (the Shannon entropy is shown as exact and asymptotic only). \textbf{(b)} Asymptotic volume exponents of the observables in (a) versus $\beta\varepsilon_\mathrm{B}$: the diversity-density exponent $\delta_D(\beta\varepsilon_\mathrm{B})-1=1/\alpha-1$, which the pool Shannon entropy per assembly shares (dotted), and the two excess-copy exponents $\delta_A(\beta\varepsilon_\mathrm{B};a_\star)$ and $\delta_A(\beta\varepsilon_\mathrm{B};1)$. Colors are matched across the two panels.}
		\label{fig:total-diversity-volume}
	\end{figure*}

	The bias $a$ tunes how strongly the measure favors long assemblies: increasing $a$ moves the observable from subextensive to marginal to superextensive growth in $V$. Figure~\ref{fig:total-diversity-volume} compares three representative complexity measures at the equal-rate point $r=1$. Panel (a) makes the tuning role explicit at three biases. The diversity density $\mathcal{D}(V)/N$ decays algebraically with $V$. The marginal observable $\mathcal{A}^{(2)}_{a_\star}(V)$, taken at $a_\star\equiv a_c^{(1)}\big|_{r=1}=-\ln(4q)\big|_{r=1}$ so that $\delta_A(a_\star)=0$, is almost scale-invariant, decreasing only through the logarithmic correction. The Sharma-style observable $\mathcal{A}^{(2)}_1(V)$ grows algebraically, because its linear weight overcompensates the rarity of long occupied assemblies. Panel (b) shows how this depends on the thermodynamic control parameter: more coagulating networks ($r\downarrow0$) push $q$ toward $1/4$, increase $\delta_A(a)$, and therefore produce faster growth of the assembly-weighted observable. This is still neutral growth in a detailed-balance network, not dynamical selection of functional structures. The temperature dependence of $\mathcal{A}^{(2)}_1$ is shown in Fig.~\ref{fig:complexity-temperature}(f), and its length spectrum at $r=1$ in panel~(c): the assembly weight $e^{a(l-1)}$ pushes weight onto longer shells, yet the sum is still controlled by the crossover scale $l_c$ (Appendix~\ref{app:assembly-weighted}), because $\mathcal{A}^{(2)}_1$ lies in the convergent range $0<\delta_A<1$.
	
	\subsection{A Unifying Scaling Law}
	\label{subsec:master-scaling}
	
	The calculations above instantiate the common shell sum of Eq.~\eqref{eq:master-sum}. Evaluating a generic summand at the crossover shell $l_c$, where $\lambda_l=O(1)$ and $4^{l_c}=(Vr)^{1/\alpha}$, gives
	\begin{equation}
		\mathcal{O}\propto \frac{(Vr)^{\delta}}{[\ln(Vr)]^{3/2}} ,
		\label{eq:master-scaling}
	\end{equation}
	where the logarithmic denominator is the Catalan signature of the crossover shell: it comes from the $l_c^{-3/2}$ prefactor in $\mathcal{N}_{l_c}$. The exponent $\delta$ is the only observable-dependent part of the leading scaling, and it is set by the length weight $w_l$. The same mechanism has been studied for more general weighted planar trees, which include the present Catalan model and would correspond here to adding an energy depending on the assembly branching~\cite{BialasBurda1996}.
	
	For a length bias $w_l\propto e^{a(l-1)}$ normalized by the assembly number $N=V\rho$,
	\begin{equation}
		\delta_A(a)=\frac{1}{\alpha}\Big(1+\frac{a}{\ln 4}\Big)-1 ,
		\label{eq:deltaA-master}
	\end{equation}
	which coincides with the exponent found above for the assembly observables. The cases obtained in this section are special values of the same law (Table~\ref{tab:complexity-scaling}). An unbiased total carries $\delta_A(0)+1=1/\alpha$ (total diversity and $S^\mathrm{Sh}_{\mathrm{pool}}$); the corresponding per-assembly density carries $\delta_A(0)=1/\alpha-1$ (the diversity density, the stochastic-sector fraction $P(L>l_c)$ of Eq.~\eqref{eq:tail-fraction-lc}, and the empirical entropy gap of Appendix~\ref{app:entropy-snapshot}); and the assembly bias $a$ tunes $\delta_A(a)$ continuously. The sign of $\delta$ classifies the growth as subextensive ($\delta<0$), marginal ($\delta=0$, logarithmic only), or superextensive ($\delta>0$). Both a larger bias $a$ and stronger coagulation ($r\downarrow0$, $\alpha\to1$) push the exponent upward.
	
	Two features of this collapse deserve emphasis, because they point beyond the present model. First, the \emph{value} of each observable trends with temperature in a measure-dependent way, and even the direction of that trend depends on the normalization. An intensive density ($1/V$, such as $S^{\mathrm{th}}/V$ or $\rho\,s^{\mathrm{Sh}}$) is $V$-independent and may rise or fall on cooling [Fig.~\ref{fig:thermodynamic-observables}(d)], whereas a per-assembly ratio ($1/N$, such as $S^{\mathrm{Sh}}_{\mathrm{pool}}/N$, $\mathcal{D}/N$, or $\mathcal{A}^{(2)}_1$) carries one power of $V$ less and drifts accordingly [Fig.~\ref{fig:complexity-temperature}(d--f)]. There is therefore no single ``complexity number'' with a universal temperature dependence. Second, and in contrast, the \emph{size-scaling exponent} $\delta$ moves the same way for every measure: cooling ($r\downarrow0$, $\alpha\to1$) raises it monotonically [Eq.~\eqref{eq:deltaA-master}]. The regimes reached still differ --- the unbiased diversity stays subextensive, its total climbing at most to extensive growth ($\propto V^{1/\alpha}$, with a logarithmic Catalan correction as $\alpha\to1$), whereas a sufficient length bias $a$ pushes $\delta_A(a)>0$ and carries the assembly observables into genuinely superextensive growth --- but the \emph{direction} of motion is shared. The exponent, not the magnitude, is the robust observable: it records that coagulation makes the neutral pool progressively more concentrated on rare, long, hierarchically deep assemblies.
	
	This points to a general principle for comparing reaction networks. The natural observable is not the instantaneous value of a structural observable but the exponent with which that observable scales in system size, together with how the exponent varies across control-parameter space: larger structural exponents mean that the network concentrates more weight on complex rare structures. Mapping $\delta(\text{parameters})$ over the phase diagram thus turns this concentration into a measurable, dimensionless quantity. In the present neutral bath the map is fixed entirely by detailed balance and the Catalan combinatorics; an irreversible or catalytic chemistry would deform these same exponents dynamically, so the neutral $\delta(\text{parameters})$ computed here provides the reference for measuring structural growth beyond neutrality.
	
	\begin{table}[t]
		\centering
		\scriptsize
		\setlength{\tabcolsep}{3pt}
		\begin{tabular}{lcccc}
			\hline
			Observable & weight $w_l$ & factor $\phi(\lambda_l)$ & norm. & exponent $\delta$ \\
			\hline
			total diversity $\mathcal{D}(V)$ & $1$ & $1-e^{-\lambda_l}$ & $1$ & $1/\alpha$ \\
			Shannon entropy $S^\mathrm{Sh}_{\mathrm{pool}}$ & $1$ & $H(\mathrm{Poi}(\lambda_l))$ & $1$ & $1/\alpha$ \\
			diversity density $\mathcal{D}(V)/N$ & $1$ & $1-e^{-\lambda_l}$ & $1/N$ & $1/\alpha-1$ \\
			assembly $\mathcal{A}^{(1,2)}_a$ & $e^{a(l-1)}$ & $\phi_{1,2}$ & $1/N$ & $\delta_A(a)$ \\
			\hline
		\end{tabular}
		\caption{The finite-size observables of this section as special cases of the scaling law~\eqref{eq:master-scaling}, $\mathcal{O}=\sum_l\mathcal{N}_l\,w_l\,\phi(\lambda_l)\propto(Vr)^{\delta}/[\ln(Vr)]^{3/2}$. The table separates the length weight $w_l$, the occupation factor $\phi(\lambda_l)$, and the normalization. Diversity and occupation Shannon entropy share $w_l=1$ and differ only through $\phi$; for assembly, $\phi_1=1-e^{-\lambda_l}$ and $\phi_2=\lambda_l-1+e^{-\lambda_l}$ correspond to $\mathcal{A}^{(1)}_a$ and $\mathcal{A}^{(2)}_a$, respectively. Normalizing by the assembly number $N=V\rho$ lowers the exponent by one, while the length bias $a$ enters through $\delta_A(a)$ of Eq.~\eqref{eq:deltaA-master}.}
		\label{tab:complexity-scaling}
	\end{table}
	
	\section{Perspectives}
	\label{sec:conclusions}
	We have studied the equilibrium organization of an unbounded chemical reaction network of assemblies, in which the number of possible species grows combinatorially with assembly size.
	The network is generated by elementary coagulation--fragmentation channels; because the assemblies are ordered binary trees, the species multiplicity has Catalan asymptotics in the assembly length, $\mathcal{N}_l\sim4^l\,l^{-3/2}$. Nevertheless, detailed balance fixes an explicit equilibrium, in which the concentration of each length-$l$ species decays geometrically, $c_l=rq^l$. In finite volume, Catalan growth competes with this decay and defines a crossover length $l_c\sim\ln V$, where the mean copy number crosses unity, separating deterministic high-copy species from a fluctuation-dominated rare sector. The deterministic sector obeys a Zipf-like rank-frequency law; the rare sector is dominated by single-copy assemblies, and in the coagulating regime its population fraction shrinks only slowly with volume, so stochasticity persists even at macroscopic volumes. In the same regime the length crossover has a temperature counterpart, a condensation temperature $T_c\sim1/\ln V$, below which a few macroscopic assemblies carry a finite fraction of the mass.
	
	This returns us to the question that opened the paper. Cayley showed that alkane isomers $\mathrm{C}_n\mathrm{H}_{2n+2}$ can be counted as bounded-valence trees whose number grows rapidly with size~\cite{Cayley1875,Polya1937}. We used the simpler Catalan family because it makes the generating function, equilibrium distribution, and crossover $l_c\sim\ln V$ explicit, but the mechanism is more general, requiring only a combinatorially growing space of structures and a conserved total mass. We therefore expect the same finite-size phenomenology in more realistic networks of this kind: a deterministic-stochastic split, rank-frequency and diversity scaling, and tunable scaling exponents. Adapting the calculation would mean replacing the Catalan count by the appropriate enumeration and adding the corresponding reaction constraints --- as in alkanes, alkenes, or other valence-constrained chemistries --- but the phenomenology would still reflect the same combinatorial--thermodynamic balance.
	
	Even the neutral reversible bath is far from trivial. Stochasticity is not confined to a microscopic correction around a deterministic limit: because the number of possible long assemblies grows combinatorially, the rare sector remains statistically important even at large volume, as reflected in the non-trivial size scaling of the diversity, the Shannon entropy, and the assembly observables. The deterministic sector, in turn, is highly structured, obeying a Zipf-like rank-frequency law in the absence of any dynamical selection mechanism. These are different aspects of the same combinatorial--thermodynamic balance, so neither sector can be treated as a correction to the other.
	
	The dynamical counterpart of this equilibrium is developed in a paper in preparation~\cite{FolenaKruszewski2026FreeExpansion}, which studies relaxation from a monomeric pool as a free expansion in structured assembly space, where entropy increases together with other complexity observables in the absence of any functional selection.
	
	The analysis of this reversible assembly dynamics is also the starting point of a broader program: understanding irreversible computational chemistries built on reversible reaction baths. Combinator-based artificial chemistries are a natural setting, since combinator assemblies are simultaneously chemical assemblies and computational objects \cite{DittrichZieglerBanzhaf2001,FontanaBuss1994,Kruszewski2022}. These are not merely formal rewriting systems: when the elementary moves are written as chemical reactions, the dynamics acquires a thermodynamic interpretation. Because the equilibrium structure described here is imprinted on the entropy and the other thermodynamic observables, studying such chemistries calls for stochastic thermodynamics, beyond the deterministic mean-field description that captures only the high-copy bulk.
	
	In an irreversible algorithmic chemistry \cite{Kruszewski2022}, this deterministic-stochastic split could become dynamically active. The deterministic sector would provide a persistent substrate, a reproducible flow of common assemblies feeding irreversible computational reactions; the rare sector, a stochastic source of occasional structures. When such a rare assembly enters an irreversible cycle, it can bias subsequent pathways, create new effective channels, or stabilize further assemblies in a history-dependent way. The neutral bath thus supplies both a deterministic background and a random exploratory mechanism, and turning on irreversible computation may convert this equilibrium hierarchy into history-dependent, non-stationary --- that is, open-ended --- dynamics.
	
	The next step is therefore not only to add irreversible reactions driving the system out of equilibrium, but to measure their thermodynamic influence. One should compare the structure generated by irreversible computational cycles against their entropy production, identify which cycles are sustained by the deterministic bath and which are seeded by rare fluctuations, and track how the accessible reaction graph changes as new assemblies become dynamically relevant. The present reversible model supplies the controlled equilibrium reference that such comparisons require.
	
	\section*{Acknowledgments}
	
	G.F. would like to thank Harold Fellermann, Raúl Toral, Emilio Hernández-García, and Antonio Fernández Peralta for insightful discussions and feedback. G.F. acknowledges funding from the Spanish Ministerio de Ciencia, Innovación y Universidades (MICIU/AEI/10.13039/ 501100011033) through the María de Maeztu project CEX2021-001164-M. Moreover, G.F. would like to thank the University of Toulouse Paul Sabatier, where this project began with financial support from the John Templeton Foundation (Grant No. 62220).
	G.K. acknowledges funding from the European Union's Horizon 2020 research and innovation program under the Marie Sk\l{}odowska-Curie Grant Agreement No. 101062294.
	The authors made use of AI large language models (Claude, Kimi) as assistive tools for coding and for reviewing the text. The authors take full responsibility for the entire content of this manuscript.
	
	\section*{Code and data availability}
	
	The data and code supporting the figures and Gillespie simulations in this work are deposited in the Zenodo collection \emph{Data and code for ``Equilibrium in a Reaction Network of Assemblies''}~\cite{FolenaKruszewski2026Zenodo}. The archived collection contains the figure-generation notebook, the processed simulation data, and the C++ Gillespie implementation used to generate the stochastic snapshots. The temporary Zenodo DOI is \href{https://doi.org/10.5281/zenodo.21269352}{10.5281/zenodo.21269352}.
	
	\bibliography{biblio}
	
	\begin{appendix}
		\begin{figure*}[ht]
			\centering
			\begin{minipage}[t]{0.31\textwidth}
				\centering
				\begin{minipage}[c][30mm][c]{\linewidth}
					\centering
					\begin{tikzpicture}[xscale=0.28,yscale=0.28,baseline=-6mm]
						\node(0)at(0.00,-6.75){$\cS$};
						\node(2)at(2.00,-6.75){$\cS$};
						\node(4)at(4.00,-4.50){$\cS$};
						\node(6)at(6.00,-4.50){$\cS$};
						\node(8)at(8.00,-4.50){$\cS$};
						\node(1)at(1.00,-4.50){\tiny$()$};
						\node(3)at(3.00,-2.25){\tiny$()$};
						\node(5)at(5.00,0.00){\tiny$()$};
						\node(7)at(7.00,-2.25){\tiny$()$};
						\draw(0)--(1);
						\draw(1)--(3);
						\draw(2)--(1);
						\draw(3)--(5);
						\draw(4)--(3);
						\draw(6)--(7);
						\draw(7)--(5);
						\draw(8)--(7);
						\node(r)at(5.00,1.69){};
						\draw(r)--(5);
					\end{tikzpicture}
				\end{minipage}
				
				\smallskip
				\textbf{Binary tree}
			\end{minipage}\hfill
			\begin{minipage}[t]{0.31\textwidth}
				\centering
				\begin{minipage}[c][30mm][c]{\linewidth}
					\centering
					\begin{forest}
						for tree={
							grow'=270,
							parent anchor=south,
							child anchor=north,
							l sep=7mm,
							s sep=4mm,
							edge+=thick
						}
						[$\cS$[$\cS$][$\cS$][$\cS$[$\cS$]]]
					\end{forest}
				\end{minipage}
				
				\smallskip
				\textbf{Planar tree}
			\end{minipage}\hfill
			\begin{minipage}[t]{0.31\textwidth}
				\centering
				\begin{minipage}[c][30mm][c]{\linewidth}
					\centering
					\begin{tikzpicture}[scale=.45,baseline=2mm]
						\draw[thick] (0,0)--(1,1)--(2,0)--(3,1)--(4,0)--(6,2)--(8,0);
						\node[below] at (0,0){$\cS$};
						\node[above] at (1,1){$\cS$};
						\node[above] at (3,1){$\cS$};
						\node[left] at (5,1){$\cS$};
						\node[above] at (6,2){$\cS$};
					\end{tikzpicture}
				\end{minipage}
				
				\smallskip
				\textbf{Dyck path}
			\end{minipage}
			\caption{Three equivalent representations of the same length-$5$ assembly, whose parenthesized expression is $\cS(\cS)(\cS)(\cS(\cS))$, in the one-atom model. The binary tree records binary coagulations; the planar tree records the ordered application structure; and the Dyck path is the contour of the planar tree. All labels are $\cS$, so only the Catalan shape distinguishes species.}
			\label{fig:assembly-representations}
		\end{figure*}
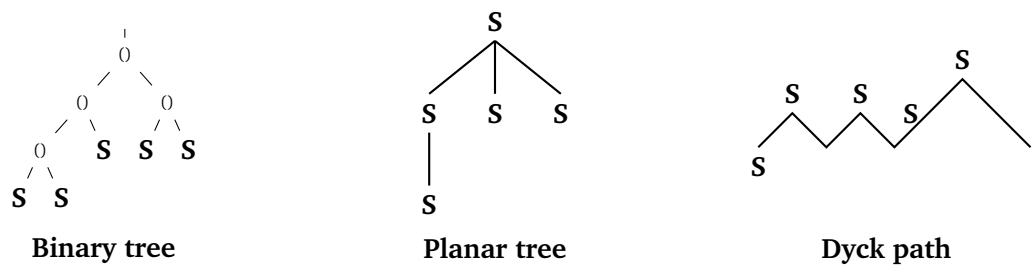
		
		\section{Equivalent Representations of Assemblies}
		\label{app:assembly-representations}
		
		An assembly in the present model is fully specified by the placement of binary coagulations among indistinguishable atoms $\cS$. Thus the same assembly can be represented in several equivalent forms~\cite[Ex.~6.19]{stanley1997enumerative}; three of them are shown here. The form used in the main text is a rooted ordered full binary tree: leaves are atoms $\cS$, and internal vertices are binary application or coagulation events. This binary-tree description is the natural one for coagulation processes based on non associative products \cite{malham_coagulation_2023}. Reading the leaves from left to right and respecting the internal vertices gives the parenthesized expression.
		
		The same object can also be represented as a rooted planar tree (Fig.~\ref{fig:assembly-representations}-center). In this representation every vertex is labeled by $\cS$, and its ordered children are the successive right arguments applied to it. Thus a planar tree with root $x$ and ordered children $y_1,\ldots,y_m$ represents the parenthesized expression $x(y_1)\cdots(y_m)$, with the same rule applied recursively to each child. Finally, the contour of this planar tree gives a Dyck path (Fig.~\ref{fig:assembly-representations}-right): each first traversal of an edge away from the root is an up step, and each return along an edge is a down step. A length-$l$ assembly has $l-1$ applications, so the corresponding Dyck path has $2(l-1)$ steps. The three descriptions are in bijection and all counted by the Catalan number $\mathcal{C}_{l-1}$~\cite{stanley1997enumerative,Dershowitz1981}.
		
		\section{Gillespie dynamics}\label{app:gillespie-sim}
		
		The simulation begins from a monomeric pool of $n_\cS = V$ copies of $\cS$, and the state at any time is represented by the multiset of active binary trees. The C++ implementation used for the stochastic snapshots is archived in the Zenodo data-and-code repository associated with this work~\cite{FolenaKruszewski2026Zenodo}. Two macroscopic observables are tracked: the total molecule count $N(t) = \sum_x n_x(t)$ and the number of fragmentable molecules $N_{\mathrm{frag}}(t) = \sum_{x:\,l(x)\ge 2} n_x(t)$, those with at least one internal bifurcation. The two reaction channels, coagulation and fragmentation, have propensities
		\begin{equation}
			a_{\mathrm{K}} = K\,N(t)\bigl(N(t)-1\bigr),
			\qquad
			a_{\mathrm{F}} = F\,N_{\mathrm{frag}}(t),
		\end{equation}
		where $K = k_K/V$ and $F = k_F$. The factor $N(t)(N(t)-1)$ counts ordered pairs of distinct molecule copies; equivalently, for a species-level reaction with $x=y$ the propensity uses $n_x(n_x-1)$ rather than $n_x^2$. Given these, the standard Gillespie procedure applies: the waiting time to the next event is drawn as
		\begin{equation}
			\Delta t = -\frac{\ln u_1}{a_{\mathrm{K}}+a_{\mathrm{F}}},
		\end{equation}
		with $u_1 \sim \mathrm{Uniform}(0,1)$, and the event type is then selected: coagulation with probability $p_\mathrm{K}=a_{\mathrm{K}}/(a_{\mathrm{K}}+a_{\mathrm{F}})$ and fragmentation with probability $p_\mathrm{F}=1-p_\mathrm{K}$.
		
		The two events act on the tree multiset as follows. A \textit{coagulation} event draws two molecules $x,y$ at random and replaces them with their join $x(y)$. A \textit{fragmentation} event draws one molecule $z=x(y)$ at random from the $N_{\mathrm{frag}}(t)$ fragmentable ones and splits it at its outermost parenthesis, replacing it with its left $x$ and right $y$ subtrees. Together these two moves implement exactly the binary-tree reaction network of Eq.~\eqref{eq:Gillespie}.
		
		\section{Asymptotic Length Scales}
		\label{app:length-scale-asymptotics}
		
		This appendix collects the asymptotic estimates underlying the finite-size length scales introduced in the main text. All estimates follow from the same two ingredients: the Catalan proliferation of possible assemblies and the exponential decay of the mean copy number of each individual assembly,
		\[
		\lambda_l=Vr q^l .
		\]
		The large-length Catalan asymptotic is
		\begin{equation}
			\mathcal{N}_l = \mathcal{C}_{l-1}
			\sim \frac{4^{l-1}}{\sqrt{\pi}\,l^{3/2}} .
			\label{eq:appendix-catalan}
		\end{equation}
		We also use
		\[
		\alpha=-\frac{\ln q}{\ln 4},
		\qquad
		\gamma=-\ln(4q),
		\]
		where $\alpha$ controls the Zipf-like rank exponent and $\gamma$ controls the far tail of the length distribution. The occupancy crossover $l_c$ is defined by $\lambda_{l_c}\sim1$, giving Eq.~\eqref{eq:lc}.
		
		\subsection{Diversity peak}
		
		The expected diversity at fixed length is
		\begin{equation}
			\langle \mathcal{D}_l \rangle
			=
			\mathcal{N}_l
			\bigl(1-e^{-Vr q^l}\bigr) ,
		\end{equation}
		so the peak occurs where the growth of $\mathcal{N}_l$ is balanced by the decrease of the occupancy factor. Treating $l$ as continuous gives the saddle condition
		\begin{equation}
			\ln 4 - \frac{3}{2l_*}
			=
			\frac{(-\ln q)\,n_*}{e^{n_*}-1},
			\qquad n_*=Vr q^{l_*}.
			\label{eq:diversity-saddle}
		\end{equation}
		For large $V$, the term $3/(2l_*)$ is subleading and $n_*$ tends to an $O(1)$ constant defined by
		\begin{equation}
			e^{n_*}=1+\alpha n_* .
		\end{equation}
		Therefore
		\begin{equation}
			l_*
			=
			l_c-\frac{\ln n_*}{-\ln q}
			+O\!\left(\frac{1}{\ln V}\right).
			\label{eq:diversity-peak}
		\end{equation}
		The explicit value of $n_*$ is not needed here. The important point is that the maximum occurs at a fixed occupancy, so the peak is shifted from $l_c$ only by an $O(1)$ amount while sharing its logarithmic growth with $V$.
		
		\subsection{Tail scales}
		
		The diversity cutoff and the largest observed length share the same tail inversion. In the rare diversity tail, $Vrq^l\ll1$, so
		\begin{equation}
			\mathcal{D}_l
			= \mathcal{N}_l\bigl(1-e^{-Vrq^l}\bigr)
			\sim \frac{Vr}{4\sqrt{\pi}}\,\frac{(4q)^l}{l^{3/2}} .
			\label{eq:appendix-Dl-tail}
		\end{equation}
		The cutoff $l_d$ is defined by $\mathcal{D}_{l_d}\sim1$.
		
		The largest observed length satisfies $V\rho\,P(L>l_{\max})\sim1$, with
		\begin{equation}
			P(L>l)=\frac{r}{\rho}\sum_{n\ge l+1}\mathcal{N}_n q^n
			\sim \frac{r q}{\rho\sqrt{\pi}(1-4q)}\,
			\frac{(4q)^l}{l^{3/2}} .
			\label{eq:appendix-tail}
		\end{equation}
		Both cases reduce to
		\begin{equation}
			\gamma l + \frac{3}{2}\ln l \sim \ln(CV),
			\qquad \gamma=-\ln(4q),
			\label{eq:appendix-tail-implicit}
		\end{equation}
		with
		\begin{equation}
			C_d=\frac{r}{4\sqrt{\pi}},
			\qquad
			C_{\max}=\frac{r q}{\sqrt{\pi}(1-4q)} .
		\end{equation}
		Inverting Eq.~\eqref{eq:appendix-tail-implicit} once gives
		\begin{equation}
			l(C)=\frac{\ln V}{\gamma}
			- \frac{3}{2\gamma}\ln\ln V
			+ \frac{1}{\gamma}\!\left[\ln C+\frac{3}{2}\ln\gamma\right]
			+ o(1) .
			\label{eq:appendix-tail-asymp}
		\end{equation}
		Thus
		\begin{equation}
			l_d=l(C_d).
			\label{eq:ld-asymp}
		\end{equation}
		\begin{equation}
			l_{\max}=l(C_{\max}).
			\label{eq:lmax-asymp}
		\end{equation}
		The two tail scales therefore share the same leading slope $1/\gamma$ and the same $-(3/2\gamma)\ln\ln V$ correction, differing only in the constant term.

		\section{Complexity-weighted Assembly}
		\label{app:assembly-weighted}
		
		The two assembly observables differ only in their occupation factor. The occupied-species version counts each realized assembly once, while the excess-copy version counts only copies beyond the first:
		\begin{equation}
			\begin{aligned}
				\mathcal{A}^{(1)}_a(V)
				&= \frac{1}{V\rho}\sum_{l\ge1} e^{a(l-1)}\mathcal{N}_l
				\bigl(1-e^{-Vrq^l}\bigr), \\
				\mathcal{A}^{(2)}_a(V)
				&= \frac{1}{V\rho}\sum_{l\ge1} e^{a(l-1)}\mathcal{N}_l
				\bigl(Vrq^l-1+e^{-Vrq^l}\bigr).
			\end{aligned}
			\label{eq:appendix-assembly-defs}
		\end{equation}
		For $a=0$, $\mathcal{A}^{(1)}_0=\mathcal{D}(V)/(V\rho)$. Expanding the occupation factor gives the shared generating-function form
		\begin{equation}
			V\rho\,\mathcal{A}^{(i)}_a(V)
			= e^{-a}\sum_{m\ge m_i}\frac{\sigma^{(i)}_m}{m!}(Vr)^m\,
			G_{\mathcal N}(e^a q^m),
			\label{eq:appendix-assembly-series}
		\end{equation}
		where $m_1=1$, $\sigma^{(1)}_m=(-1)^{m+1}$ for the occupied-species observable, and $m_2=2$, $\sigma^{(2)}_m=(-1)^m$ for the excess-copy observable. The convergence boundary is set by the first nonzero term, because $G_{\mathcal N}(u)$ has radius of convergence $1/4$. Hence
		\begin{equation}
			a_c^{(1)}=-\ln(4q),
			\qquad
			a_c^{(2)}=-\ln(4q^2)=a_c^{(1)}-\ln q .
		\end{equation}
		The excess-copy observable is finite over the larger interval because the $m=1$ term has been removed.
		
		For both definitions, the saturated side and rare tail of the crossover have the same leading volume dependence. The normalized observables scale as
		\begin{equation}
			\mathcal{A}^{(1)}_a(V),\ \mathcal{A}^{(2)}_a(V)
			\propto
			\frac{(Vr)^{\delta_A(a)}}{[\ln(Vr)]^{3/2}},
			\qquad
			\delta_A(a)=\frac{a-a_c^{(1)}}{-\ln q}.
			\label{eq:appendix-assembly-asymptotic}
		\end{equation}
		
		The location of the dominant shell follows from the same rescaling $\hat l=l/l_c$ used for diversity [Eq.~\eqref{eq:diversity-tent}]. Writing the summand as $e^{\ln(Vr)\,\psi_A(\hat l)}$, the two observables have
		\begin{equation}
			\psi^{(1)}_A(\hat l)=
			\begin{cases}
				(1+\delta_A)\,\hat l, & \hat l\le1,\\[2pt]
				1+\delta_A\,\hat l, & \hat l\ge1,
			\end{cases}
			\qquad
			\psi^{(2)}_A(\hat l)=
			\begin{cases}
				1+\delta_A\,\hat l, & \hat l\le1,\\[2pt]
				2+(\delta_A-1)\,\hat l, & \hat l\ge1.
			\end{cases}
			\label{eq:assembly-tent}
		\end{equation}
		Both rate functions have their crossover-shell value at $\hat l=1$. This point is the dominant shell only when the left branch rises and the right branch falls: for $\mathcal{A}^{(1)}$ this requires $-1<\delta_A<0$ (i.e. $-\ln4<a<a_c^{(1)}$), a subrange of its convergence domain $a<a_c^{(1)}$ that contains the diversity point $a=0$; for $\mathcal{A}^{(2)}$ it requires $0<\delta_A<1$, i.e. $a_c^{(1)}<a<a_c^{(2)}$. In these ranges the bias shifts the peak only by an $O(1)$ amount and leaves the scale $l_c\sim\ln(Vr)$ unchanged. Outside them the peak detaches from $l_c$: for $\mathcal{A}^{(2)}$ at $\delta_A<0$ it moves to short high-occupancy shells, while as $\delta_A\to1$ it moves toward the longest realized assemblies.
		
		This is why the linear assembly choice $a=1$ is admissible for $\mathcal{A}^{(2)}_a$ even when it lies beyond the finite window of $\mathcal{A}^{(1)}_a$. The sign of
		\begin{equation}
			\delta_A(\varepsilon_\mathrm{B};a)=\frac{1}{\alpha(\varepsilon_\mathrm{B})}-\frac{a}{\ln q(\varepsilon_\mathrm{B})}-1
		\end{equation}
		separates decaying, marginal, and growing normalized assembly observables (we are considering $T=1$ and varying $\varepsilon_\mathrm{B}$). Multiplication by $V\rho$ shifts these exponents by one for the extensive numerators. For the unbiased density observable one recovers the diversity-density exponent $\delta_A(\varepsilon_\mathrm{B};0)=1/\alpha-1$, as illustrated in Fig.~\ref{fig:total-diversity-volume}.

		\section{Subextensive Shannon Pool Entropy}
		\label{app:shannon_pool}
		
		Start with the entropy of a single extracted assembly. At equilibrium, the total density
		\begin{equation}
			\rho = \sum_x c_x = \sum_x e^{-\beta\varphi(x)} ,
		\end{equation}
		plays the role of the one-body partition function, with normalized probability $p_x=c_x/\rho$.
		The Shannon entropy of this assembly distribution is
		\begin{equation}
			\begin{aligned}
				s^\mathrm{Sh}
				&\equiv -\sum_x p_x \ln p_x \\
				&= \ln \rho +\beta \sum_x p_x \varphi(x)\\
				&= (1-\beta\,\partial_\beta) \ln\rho \equiv s^\mathrm{th}_{\mathrm{one}} \ .
			\end{aligned}
			\label{eq:spop-q-derivative}
		\end{equation}
		
		In the last line we recover the thermodynamic entropy of the one-body extracted-assembly ensemble by differentiating the partition function $\ln\rho$ with the operator $\mathcal{O}_\beta \equiv 1-\beta\,\partial_\beta$. For this one-body ensemble, the Shannon entropy $s^\mathrm{Sh}$ and the thermodynamic entropy $s^\mathrm{th}_{\mathrm{one}}$ are equivalent. This equivalence does not extend automatically to the whole pool.
		
		For the whole pool, the microscopic state is the occupation vector $\mathbf{n}=\{n_x\}$. Writing $N=V\rho$ for the mean total number of assemblies, the normalized multivariate Poisson law is
		\begin{equation}
			\mathbb{P}_{\mathbf{n}}
			= \prod_x e^{-\lambda_x}\frac{{\lambda_x}^{n_x}}{n_x!} \qquad
			\text{with means}	\quad \lambda_x = N p_x = V c_x = \langle n_x \rangle
		\end{equation}
		The Shannon entropy of this occupation-number distribution is
		\begin{equation}
			\begin{aligned}
				S^\mathrm{Sh}_{\mathrm{pool}}
				&\equiv -\sum_{\mathbf n} \mathbb{P}_{\mathbf n} \ln \mathbb{P}_{\mathbf n} \\
				&= -\sum_x \sum_{n_x\geq0} e^{-\lambda_x}\frac{\lambda_x^{n_x}}{n_x!} \ln \left (e^{-\lambda_x}\frac{\lambda_x^{n_x}}{n_x!}\right ) \\
				&= \sum_x \left [ \lambda_x - \langle n_x \rangle \ln \lambda_x + \langle \ln(n_x !) \rangle \right ]
			\end{aligned}
			\label{eq:entropy-sh-pool}
		\end{equation}
		where $\langle A \rangle = \sum_\mathbf{n} \mathbb{P}_{\mathbf{n}} A_\mathbf{n}$ denotes expectation over $\mathbb{P}_{\mathbf{n}}$. The second line follows by additivity of Shannon entropy for independent factors.\footnote{Given $P(n_1,n_2)=P(n_1)P(n_2)$, with $\sum_{n_1}P_{n_1}=1$ and $\sum_{n_2}P_{n_2}=1$, we have that $\sum_{n_1,n_2}P(n_1,n_2)\ln P(n_1,n_2) = \sum_{n_1}P_{n_1}\ln P_{n_1}+\sum_{n_2}P_{n_2}\ln P_{n_2}$.}
		
		We now return to the thermodynamic pool entropy introduced above. The pool partition function includes the indistinguishability of identical assemblies through the factorial weights,
		\begin{equation}
			Z_{\mathrm{pool}}
			= \sum_{\mathbf{n}} \prod_x \frac{(V c_x)^{n_x}}{n_x!}
			= \exp\!\left(V\sum_x c_x\right)
			= e^{V\rho} .
			\label{eq:entropy-Z}
		\end{equation}
		Applying $\mathcal{O}_\beta$ to $\ln Z_{\mathrm{pool}}=V\rho$ reproduces the thermodynamic pool entropy in Eq.~\eqref{eq:entropy-sum-species},
		\begin{equation}
			\begin{aligned}
				S^\mathrm{th}_{\mathrm{pool}}
				&\equiv \mathcal{O}_\beta \ln Z_{\mathrm{pool}}
				= V(\rho-\beta\partial_\beta \rho)
				= \sum_x \left [ V c_x - \langle n_x \rangle \ln c_x \right ] \\
				&= -\sum_{\mathbf n} \mathbb{P}_{\mathbf n} \left[ \ln \mathbb{P}_{\mathbf n} -\sum_x \ln(\frac{V^{n_x}}{n_x!}) \right] = S^\mathrm{Sh}_{\mathrm{pool}} + S_\mathrm{im} ,
			\end{aligned}
			\label{eq:entropy-pool}
		\end{equation}
		where $S_\mathrm{im}= \sum_{\mathbf n} \mathbb{P}_{\mathbf n} \sum_x \ln(\frac{V^{n_x}}{n_x!})$ is the ideal-mixture contribution associated with placing indistinguishable copies of each assembly in the volume. The thermodynamic and Shannon entropies of the pool are therefore not equivalent~\cite{SchmiedlSeifert2007}, and the two terms of Eq.~\eqref{eq:entropy-pool} play very different roles. The ideal-mixture term $S_\mathrm{im}$ is extensive, while the Shannon term $S^\mathrm{Sh}_{\mathrm{pool}}$ is controlled by the same rare crossover shell around $l_c$ that sets the occupied diversity (Sec.~\ref{subsec:diversity-complexity}) and is therefore subextensive. In the large-volume limit the extensive pool entropy is thus carried entirely by the ideal-mixture term,
		\begin{equation}
			S^\mathrm{th}_{\mathrm{pool}}\simeq S_\mathrm{im},
			\qquad
			\frac{S^\mathrm{Sh}_{\mathrm{pool}}}{S^\mathrm{th}_{\mathrm{pool}}}\xrightarrow[V\to\infty]{}0 ,
			\label{eq:sh-pool-subextensive}
		\end{equation}
		while the occupation-number Shannon entropy survives only as a slowly vanishing correction. This ratio decays as $V^{-(\alpha-1)/\alpha}$ up to logarithmic factors, with exactly the exponent of the stochastic-sector fraction $P(L>l_c)$ of Eq.~\eqref{eq:tail-fraction-lc}. Both are set by the same rare crossover shell, and at $r=1$ the exponent is only $(\alpha-1)/\alpha\approx0.04$: the rare sector of assembly space, though thermodynamically negligible, leaves a long-lived imprint on the information-like Shannon entropy of the pool.
		
		This subextensive scaling follows from the same rescaled-length saddle estimate used in the main text for diversity. At fixed thermodynamic parameters,
		\begin{equation}
			S^\mathrm{Sh}_{\mathrm{pool}}
			= \sum_l \mathcal{N}_l\,H\bigl(\mathrm{Poi}(\lambda_l)\bigr),
			\qquad
			\lambda_l=Vrq^l,
		\end{equation}
		with $q=4^{-\alpha}$ and $l_c=\ln(Vr)/(\alpha\ln4)$. Setting $\hat l=l/l_c$ gives $\lambda_l=(Vr)^{1-\hat l}$ and $4^l=(Vr)^{\hat l/\alpha}$. The entropy of a Poisson variable has the limiting forms
		\begin{equation}
			H\bigl(\mathrm{Poi}(\lambda)\bigr)
			\simeq \tfrac12\ln(2\pi e\lambda)\quad(\lambda\gg1),
			\qquad
			H\bigl(\mathrm{Poi}(\lambda)\bigr)
			\simeq \lambda(1-\ln\lambda)\quad(\lambda\ll1).
		\end{equation}
		These logarithmic factors change only subexponential prefactors. The leading exponential dependence of the shell contribution is therefore
		\begin{equation}
			\mathcal{N}_l\,H\bigl(\mathrm{Poi}(\lambda_l)\bigr)
			\sim
			\frac{\exp\!\left[\ln(Vr)\,\psi_{\mathrm{Sh}}(\hat l)\right]}{[\ln(Vr)]^{3/2}},
			\qquad
			\psi_{\mathrm{Sh}}(\hat l)=
			\begin{cases}
				\hat l/\alpha, & \hat l\le1,\\[2pt]
				1-\hat l(\alpha-1)/\alpha, & \hat l\ge1.
			\end{cases}
		\end{equation}
		The rate function peaks at the crossover shell $\hat l=1$, where $\lambda_l=O(1)$ and $\psi_{\mathrm{Sh}}(1)=1/\alpha$. Thus the sum is controlled by lengths $l=l_c+O(1)$ and
		\begin{equation}
			S^\mathrm{Sh}_{\mathrm{pool}}
			\propto
			\frac{(Vr)^{1/\alpha}}{[\ln(Vr)]^{3/2}},
		\end{equation}
		as stated in Sec.~\ref{subsec:entropy}.
		
		\section{Finite-Volume Snapshot Entropy}
		\label{app:entropy-snapshot}
		
		This appendix isolates three points behind the slow convergence of the empirical entropy estimator in Sec.~\ref{subsec:entropy}: what is measured in a finite snapshot, how the finite-volume bias can be written as a shell sum, and why that bias has the same scaling as the pool occupation-number entropy.
		
		\subsection{Snapshot Estimators}
		
\begin{figure}[t]
			\centering
			\includegraphics[width=\columnwidth]{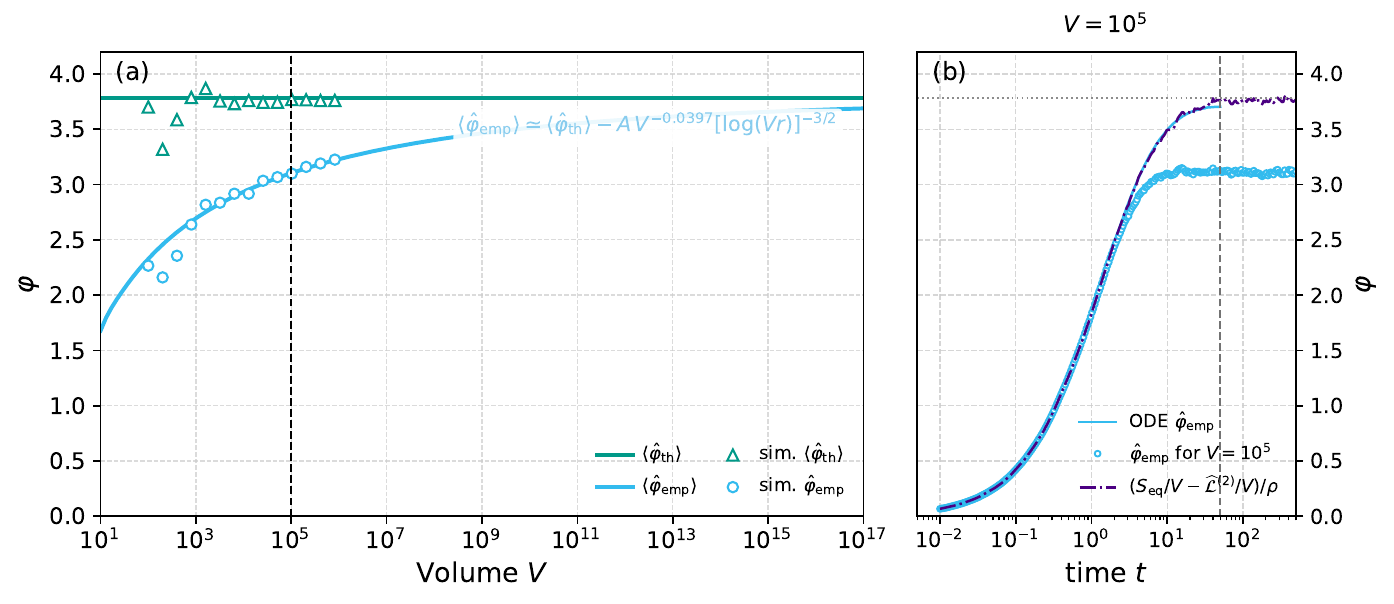}
			\caption{Single-snapshot one-copy free energy estimators and their time evolution at $r=1$. \textbf{(a)} Estimators versus volume. Green triangles show $\widehat{\varphi}_\mathrm{th}$, while the green horizontal line is its exact expectation $\langle\varphi\rangle$. Cyan circles show $\widehat{\varphi}_\mathrm{emp}$, and the cyan solid curve is its finite-volume expectation. The volume points are taken from the Gillespie snapshot at time $t=50$. \textbf{(b)} Log-time evolution of the empirical estimator $\widehat{\varphi}_\mathrm{emp}$. Simulation snapshots are shown as markers only, while the continuous ODE references are obtained by the cutoff-extrapolation procedure described in a paper in preparation~\cite{FolenaKruszewski2026FreeExpansion}. The vertical dashed line marks $t=50$. The strongly converging estimator (purple dashed-dotted line) is obtained from a stochastic Lyapunov functional analyzed there.}
			\label{fig:entropy-volume}
		\end{figure}

		Two snapshot estimators of the one-copy free energy are compared in Fig.~\ref{fig:entropy-volume}. When the equilibrium one-copy free energy is known, the model-based estimator
		\begin{equation}
			\beta\widehat{\varphi}_\mathrm{th}(\mathbf{n})=
			\sum_x\frac{n_x}{N}\,\beta\varphi(x)
			=-
			\ln r-\ln q\sum_x\frac{n_x}{N}\,l(x)
			\label{eq:tilde-phi-estimator}
		\end{equation}
		depends, by mass conservation $\sum_x n_x l(x)=V$, only on the mean length $V/N$ and is therefore strongly self-averaging. The model-free plug-in estimator
		\begin{equation}
			\beta\widehat{\varphi}_\mathrm{emp}(\mathbf{n})=
			-\sum_{x:\,n_x>0}\frac{n_x}{N}\,
			\ln\!\left(\frac{n_x}{V}\right)
			=\widehat{s}^{\mathrm{Sh}}(\mathbf{n})-\ln\rho
			\label{eq:hat-phi-estimator}
		\end{equation}
		[Eq.~\eqref{eq:entropy-snapshot-estimator}] instead reads probabilities from the sample itself. It is biased because rare assemblies may be absent and visible assemblies have fluctuating integer copy numbers.
		
		\subsection{Systematic Bias}
		
		To evaluate the expectation of the snapshot Shannon entropy, we average over the Poisson equilibrium representation and replace the random denominator $N$ in the plug-in estimator by its mean $V\rho$. This changes finite-size prefactors but not the leading crossover-shell exponent. For one assembly of length $l$,
		\begin{equation}
			p_l=\frac{c_l}{\rho}=\frac{\lambda_l}{V\rho},
			\qquad
			\lambda_l=Vc_l=Vrq^l .
		\end{equation}
		In a finite snapshot the observed copy number is $k\sim\mathrm{Poi}(\lambda_l)$, with the $k=0$ contribution set to zero. Averaging the plug-in contribution over $k$ and summing over length shells gives
		\begin{equation}
			s^{\mathrm{Sh}}(V)
			= -\sum_{l\geq 1}\mathcal{N}_l\,
			\mathbb{E}_{\mathrm{Poi}(\lambda_l)}
			\!\left[\frac{k}{V\rho}\ln\!\left(\frac{k}{V\rho}\right)\right],
			\label{eq:entropy-truncated}
		\end{equation}
		where $s^{\mathrm{Sh}}(V)=\mathbb{E}[\widehat{s}^{\mathrm{Sh}}]$ denotes the expected snapshot entropy. Subtracting this from the infinite-volume entropy gives
		\begin{equation}
			s^{\mathrm{Sh}}-s^{\mathrm{Sh}}(V)
			=
			\frac{1}{V\rho}
			\sum_{l\geq1}\mathcal{N}_l\,\Phi(\lambda_l),
			\qquad
			\Phi(\lambda)=\mathbb{E}_{\mathrm{Poi}(\lambda)}[k\ln k]-\lambda\ln\lambda .
			\label{eq:entropy-gap-exact}
		\end{equation}
		The function $\Phi$ contains both sources of bias: missing support from species with $k=0$ and finite-copy fluctuations among species with $k>0$. Its limits are
		\begin{equation}
			\Phi(\lambda)=\frac12+O(\lambda^{-1})\quad(\lambda\gg1),
			\qquad
			\Phi(\lambda)=-\lambda\ln\lambda+O(\lambda^2)\quad(\lambda\ll1),
		\end{equation}
		The dominant contribution therefore comes from the crossover shell $\lambda_l=O(1)$, where the entropy-bias factor remains bounded. 
		
		\subsection{Scaling and Pool Entropy}
		
		Equation~\eqref{eq:entropy-gap-exact} is a per-assembly shell sum. The master scaling law therefore gives
		\begin{equation}
			s^{\mathrm{Sh}}-s^{\mathrm{Sh}}(V)
			\sim
			A(r,q)\,V^{-(\alpha-1)/\alpha}\,[\ln(Vr)]^{-3/2},
			\label{eq:entropy-asymptotic}
		\end{equation}
		with an $O(1)$ prefactor $A(r,q)$. The exponent is the same as for the diversity density and the stochastic-sector fraction. At $r=1$, $(\alpha-1)/\alpha\approx0.04$, so the empirical entropy is only weakly self-averaging.
		
		The pool occupation-number entropy is controlled by the same shell. Comparing Eq.~\eqref{eq:entropy-gap-exact} with Eq.~\eqref{eq:entropy-sh-pool} gives
		\begin{equation}
			S^{\mathrm{Sh}}_{\mathrm{pool}}
			= N\,\big[\,s^{\mathrm{Sh}}-s^{\mathrm{Sh}}(V)\,\big] + R_V,
			\label{eq:pool-equals-gap}
		\end{equation}
		where $R_V$ is a residual finite-copy contribution. This residual changes the prefactor, so the relation consists in a shared leading scale an not an exact identity of amplitudes:
		\begin{equation}
			s^{\mathrm{Sh}}-s^{\mathrm{Sh}}(V)
			\sim \frac{S^{\mathrm{Sh}}_{\mathrm{pool}}}{N},
			\qquad
			S^{\mathrm{Sh}}_{\mathrm{pool}}
			\propto \frac{(Vr)^{1/\alpha}}{[\ln(Vr)]^{3/2}} .
		\end{equation}
		Thus the bias of the naive snapshot estimator is the per-assembly imprint of the same rare occupation-number entropy that makes $S^{\mathrm{Sh}}_{\mathrm{pool}}$ subextensive.
		
	\end{appendix}
	
\end{document}